\documentclass{aa}
\usepackage[utf8]{inputenc}
\usepackage{graphicx}   
\usepackage{amssymb}
\usepackage[fleqn]{amsmath}
\usepackage[varg]{txfonts}
\usepackage{epstopdf}
\usepackage{multirow}
\usepackage{graphicx}
\usepackage{expdlist}
\usepackage{amssymb}
\usepackage{natbib}
\usepackage{color}
\usepackage{epsfig}
\usepackage{pdflscape}
\usepackage{lscape}
\usepackage{rotating}
\usepackage{supertabular}
\usepackage{lscape}
\usepackage{verbatim}
\usepackage{threeparttable}
\usepackage{mathrsfs}
\usepackage{pdflscape}
\usepackage{longtable}
\usepackage[]{caption}
\usepackage{threeparttablex} % for "ThreePartTable" environment
\usepackage{booktabs}        % for well-spaced horizontal rules

\newcommand{\mum}{$\mu$m }
\newcommand{\eg}{e.g.,}
\newcommand{\ie} {i.e.,}
\newcommand{\kms}{~km~s$^{-1}$}
\newcommand{\mo} {M$_{\odot}$}

\newcommand{\jybeam}{Jy\,beam$^{-1}$}
\newcommand{\lsun}{L$_{\odot}$}
\newcommand{\msun}{M$_{\odot}$}
\newcommand{\rah}{$^{\mbox{\scriptsize h}}$}
\newcommand{\ram}{$^{\mbox{\scriptsize m}}$}
\newcommand{\ras}{$^{\mbox{\scriptsize s}}$}
\newcommand{\decd}{$^{\circ}$}
\newcommand{\decm}{$'$}
\newcommand{\decs}{$\farcs$}

\newcommand{\phn} {\phantom{0}}
\title{Role of the magnetic field in the fragmentation process: the case of G14.225-0.506}

\titlerunning{Magnetic fields in the IRDC G14.2}

\author{N. A\~nez-L\'opez \inst{1,2} \and G. Busquet \inst{1,2} \and P.~M. Koch \inst{3} \and J.~M. Girart \inst{1,2} \and H. B. Liu\inst{3} \and F. Santos\inst{4} \and N.~L. Chapman\inst{5} \and G. Novak\inst{5} \and A. Palau\inst{6} \and P.~T.~P. Ho\inst{3} 
\and Q. Zhang\inst{7}}

\institute{Institut de Ci\`encies de l'Espai (ICE, CSIC), Can Magrans s/n, E-08193 Cerdanyola del Vall\`es, Catalonia
\and 
Institut d'Estudis Espacials de catalunya (IEEC), E-08034, Barcelona, Catalonia
\and 
Academia Sinica, Institute of Astronomy and Astrophysics, Taipei 10617, Taiwan % pmkch@asiaa.sinica.edu.tw 
\and
Max-Planck-Institute for Astronomy, K\"{o}nigstuhl 17, 69117 Heidelberg, Germany
\and 
Center for Interdisciplinary Exploration and Research in Astrophysics (CIERA), and Department of Physics \& Astronomy, Northwestern University, 2145 Sheridan Rd, Evanston, IL, 60208, USA
\and
Instituto de Radioastronom\'ia y Astrof\'isica, Universidad Nacional Aut\'onoma de M\'exico, P.O. Box 3-72, 58090, Morelia, Michoac\'an, M\'exico
\and
Harvard-Smithsonian Center for Astrophysics, Cambridge, MA, 02138, USA}
\authorrunning{A\~nez-L\'opez et al.}

\date{ Received {\it date} / Accepted {\it date} }

\abstract{Magnetic fields are predicted to play a significant role in the formation of filamentary structures and their fragmentation to form stars and star clusters.}
{We aim to investigate the role of the magnetic field in the process of core fragmentation toward the two hub--filament systems in the infrared dark cloud G14.225-0.506, which present different levels of fragmentation.}
{We performed observations of the thermal dust polarization at 350~\mum using the Caltech Submillimeter Observatory (CSO) with an angular resolution of 10$''$ toward the two hubs (Hub-N and Hub-S) in the infrared dark cloud G14.225-0.506. We additionally applied the polarization--intensity-gradient method to estimate the significance of the magnetic field over the gravitational force.}
{The sky-projected magnetic field in Hub-N shows a rather uniform structure along the east--west orientation, which is roughly perpendicular to the major axis of the hub--filament system. The intensity gradient in Hub-N displays a single local minimum coinciding with the dust core MM1a detected with interferometric observations.
Such a prevailing magnetic field orientation is slightly perturbed when approaching the dust core.
Unlike the northern Hub, Hub-S shows two local minima, reflecting the bimodal distribution of the magnetic field. 
In Hub-N, both east and west of the hub--filament system, the intensity gradient and the magnetic field are parallel whereas they tend to be perpendicular when penetrating the dense filaments and hub. Analysis of the $|\delta|$- and $\Sigma_{B}$-maps indicates that, in general, the magnetic field cannot prevent gravitational collapse,
both east and west, suggesting that the magnetic field is initially dragged by the infalling motion and aligned with it, or is channeling material toward the central ridge from both sides. 
Values of $\Sigma_{B}\gtrsim1$ are found toward a north--south ridge encompassing the dust emission peak, indicating that in this region magnetic field dominates over gravity force, or that with the current angular resolution we cannot resolve a hypothetically more complex structure. 
We estimated the magnetic field strength, the mass-to-flux ratio, and the Alfvén Mach number, and found differences between the two hubs.}
{The different levels of fragmentation observed in these two hubs could arise from differences in the  properties of the magnetic field rather than from differences in the intensity of the gravitational field because the density in the two hubs is similar. However, environmental effects could also play a role.}


\usepackage[normalem]{ulem}

\begin{document}

\maketitle

%%%%%%%%%%%%%%%%%%%%%%%%%%%%%%%%%%%%%%%%%%%%%%%%%%%%%%%%%%%%%%%%%%%%%%%
\section{Introduction}
%%%%%%%%%%%%%%%%%%%%%%%%%%%%%%%%%%%%%%%%%%%%%%%%%%%%%%%%%%%%%%%%%%%%%%%

Recent observations show that filaments are prevailing structures in molecular clouds
%\LEt{ please alter the coding so that "e.g.," has no spaces.} 
\citep[\eg][]{Myers2009,Molinari2010,Andre2014,Rivera-Ingraham2016,Dhabal2018}, but their formation mechanism and the effect of interplay between gravity, turbulence, and magnetic fields on the origin and evolution of filamentary structures remains unclear, and their specific role in the fragmentation process to form star clusters is still under debate. 
Several models have been proposed to explain the formation and evolution of filamentary
structures, such as the magnetized filament \citep{Inoue&Fukui2013, VanLoo2014}. Others consider turbulent environments \citep{Padoan2001, Moeckel2015} and models which consider both turbulence and magnetic fields \citep{Kirk2015, Federrath2016}.
One possible scenario is Global, Hierarchical Gravitational collapse (GHC;  \citealt{GomezandVazque-Semadeni2014, Smith2014,Smith2016,Vazquez-Semadeni2019}), according to which all scales accrete material from their parent structures, and filamentary accretion flows form as a natural consequence of the gravitational collapse of a highly inhomogeneous cloud.

Magnetic fields are believed to be important at cloud and core scales \citep[e.g.,][]{Girart2009, Zhang2014, Li2015, Beltran2019}. In particular, observations suggest that magnetic fields in molecular clouds are perpendicular to the filaments \citep[\eg][]{Sugitani2011,Palmeirim2013,Pillai2015,Santos2016,PlanckColl2016}. There is also observational evidence of gas flows along the filaments that converge into the hubs \citep[\eg][]{Kirk2013,Peretto2014,Williams2018,Chen2019}. 

The infrared dark cloud G14.225-0.506 (hereafter IRDC G14.2), located at a distance of 1.98~kpc \citep{Xu2011}, is part of the extended ($77\times15$~pc) and massive ($>10^5$~\mo) molecular cloud discovered by \citet{Elmegreen1976}, located southwest of the Galactic \ion{H}{II} region M17. 
Observations of the dense gas emission reveal a network of filaments constituting two hub--filament systems \citep{Busquet2013,Chen2019}. The hubs are associated with a rich population of protostars and young stellar objects \citep{PovichWhitney2010,Povich2016}, and appear more compact, warmer ($T_{\mathrm{rot}}\simeq$15~K), and show large velocity dispersion and larger masses per unit length than the surrounding dense filaments, suggesting they are the main sites of stellar activity within the cloud \citep{Busquet2013}.

The sky-projected magnetic field morphology in G14.2 has been mapped through interstellar polarization of background starlight at optical and near-infrared wavelengths \citep{Santos2016}. At large scales, \citet{Santos2016} find that the magnetic field in G14.2 is tightly perpendicular to the molecular cloud and to the dense filaments. The magnetic field strength was estimated using the Davis-Chandrasekhar-Fermi method to be in the range of $320-550$~$\mu$G, 
which lead to prevailing sub-alfv\'enic conditions, although the field strength does not seem to be sufficient to prevent the gravitational collapse of the hubs and filaments. These general features are suggestive of a scenario in which the magnetic fields played a significant role in regulating gravitational collapse on the $\sim$30 to $\sim$2~pc scale range.

Focusing on the hubs in G14.2, high-angular resolution observations ($\sim1\farcs$5) carried out with the Submillimeter Array (SMA) reveal that the two hubs are physically indistinguishable at 0.1~pc scales, despite clearly presenting a different level of fragmentation when observed with a spatial resolution of 0.03~pc: Hub-S being more fragmented than Hub-N \citep{Busquet2016}. Despite the presence of a bright 
%\LEt{Please spell out all acronyms the first time they appear in the paper, followed by the abbreviation in parentheses, both in the abstract and again in the main text. After that, please only use the abbreviation. See A and A language guide Section 5.2.4 www.aanda.org/language-editing}
Infrared Astronomical Satellite (IRAS) source ($L\simeq10^4$~\lsun) close to Hub-N, \citet{Busquet2016} show that all derived physical properties such as the density and temperature profiles, the level of turbulence,
the magnetic field around the hubs, and the rotational-to-gravitational energy are remarkably similar in both hubs. The authors conclude that the lower fragmentation level observed in Hub-N could result from the effects of UV radiation from the nearby H\ion{}{II} region, evolutionary effects, and/or stronger magnetic field inside the hubs than originally derived.

The polarization of dust thermal emission at submillimeter wavelengths provides a method to study magnetic field properties \citep{Hildebrand2000}. It is thought that dust grains are aligned with their longer axis perpendicular to the magnetic field lines; thus, the emitted light appears to be polarized perpendicular to the field lines \citep[see][for reviews of grain-alignment mechanisms]{Lazarian2007,Andersson2015}. Therefore, dust polarization measurements probe the plane of sky-projected magnetic field structure in dense regions.

In this paper, we present CSO 350~\mum polarimetric observations toward the two hubs in the IRDC G14.2. The paper is organized as follows.
Section~\ref{s:obs} describes the observations and data-reduction process.
In Section~\ref{s:res} we present the main results, and we analyze the thermal dust polarized emission in 
Section~\ref{sec:ana}.
Finally, in Section~\ref{s:discuss} we discuss our findings and list our main conclusions in Section~\ref{s:concl}. 

%%%%%%%%%%%%%%%%%%%%%%%%%%%%%%%%%%%%%%%%%%%%%%%%%%%%%%%%%%%%%%%%%%%%%%%
\section{Observations and data reduction}\label{s:obs}
%%%%%%%%%%%%%%%%%%%%%%%%%%%%%%%%%%%%%%%%%%%%%%%%%%%%%%%%%%%%%%%%%%%%%%%

SHARP \citep{Li2008} is a fore-optics module  that adds polarimetric capabilities to SHARC-II, a $12\times32$ pixel bolometer array \citep{Dowell2003}. SHARP separates the incident radiation into two orthogonal polarization states that are then imaged side-by-side on the SHARC-II array. SHARP includes a half-wave plate (HWP) located upstream from the polarizing splitting optics. Polarimetric observations with SHARP involve carrying out chop-nod photometry at each of four HWP rotation angles. 
Observations carried out with SHARP prior to December 2011 suffered correlated noise, the treatment of which required special techniques \citep{Chapman2013}. The techniques used included $\chi^2$ analysis followed by error inflating, or the use of generalized Gauss-Markov theorem. In December 2011, the ``cold load mirrors'' in SHARP \citep{Li2008} were replaced with warm absorbers; and subsequent testing revealed that this increased the photon noise but largely eliminated the correlated noise problem. Accordingly, no treatment for correlated noise was applied to the present data set.

The wavelength of observation was 350~\mum and the effective beam size was $\sim10''$. We observed the IRDC G14.2 with 30 pointings to cover the entire cloud complex 
(see Appendix, Fig.~\ref{fig:regions} and Table~\ref{tab:pointings}).
Observations were obtained remotely on June 09, 10, 11, 12, 14, and 17 of 2015.
On each night, we observed the young stellar object IRAS~16293-2422 for initial focusing and pointing calibrations, and observed Neptune at the end of the operation for absolute flux calibration.
We employed a 210$''$ chop throw for calibration observations, and employed a 300$''$ chop throw for target source observations.

The data were calibrated using the method presented in \citet{Davidson2011} and \citet{Chapman2013}. 
The polarization intensity shown in the present paper has been debiased.
SHARP data analysis is carried out in two steps. In the first step, each individual half-wave plate cycle is processed to obtain six 12$\times$12 pixel maps, one for each of the Stokes parameters I, Q, U, and one for the corresponding error maps. 
Uncertainties are computed using the variance in the individual total and polarized flux measurements \citep{Dowell1998, Hildebrand2000}. In the second step of the data analysis, single-cycle maps are combined to form the final maps and the errors are propagated to the final maps.
The sky rotation is taken into account by interpolating the single-cycle maps onto a regular equatorial-coordinate grid \citep{HoudeVaillancourt2007}.
Corrections for changing atmospheric opacity as well as for instrumental polarization and polarimetric efficiency are also made during this second analysis step \citep{Kirby2005,Li2008}. 
The percentage and position angle of polarization and their associated errors are computed for each sky position through standard techniques \citep[see][]{Hildebrand2000}. 

A common problem of the chop-node mode technique arises when the emission of the reference (off-source) position is comparable in intensity to the emission from the target source. 
After careful inspection of the emission in the off-source positions using the 350~\mum\ dust continuum map \citep{Busquet2016,Lin2017}, we flagged scans in which the emission coming from the off-source position was larger, on average, than $\sim$20$\%$ of the intensity peak in the target source position. 
Following these criteria, we flagged 10 cycles toward
Hub-N and 90 toward Hub-S, $\sim$ 11$\%$, and $\sim$ 42$\%$ of the total, respectively.

We used Matplotlib\footnote{https://docs.scipy.org} to calculate the histogram of polarization angles. This software allows the user to select different methods to calculate the optimal bin width and consequently the number of bins. While with the Freedman Diaconis (FD) estimator  the bin width is proportional to the interquartile range and inversely proportional to the cube root of \textbf{a}, where \textbf{a} is the number of data points, with the Sturges Estimator the number of bins is the base 2\,log of \textbf{a}. In this work, we used the `auto' mode which selects 
the Sturges value for small datasets, while for larger datasets will usually default to FD which
%\textcolor[rgb]{0.984314,0.00784314,0.027451}{maximum of the two} \LEt{ the meaning here is unclear; please consider rewording or expanding slightly.}estimators, Sturges and FD, and 
avoids the overly conservative behavior of FD and Sturges for small and large data sets, respectively. 
We note that because of the low sample size for polarization detections in Hub-S, the histogram of the polarization angles was built considering a lower number of bins.

%%%%%%%%%%%%%%%%%%%%%%%%%%%%%%%%%%%%%%%%%%%%%%%%%%%%%%%%%%%%%%%%%%%%%%%
\section{Results}\label{s:res}
%%%%%%%%%%%%%%%%%%%%%%%%%%%%%%%%%%%%%%%%%%%%%%%%%%%%%%%%%%%%%%%%%%%%%%%

\begin{figure}[!t]
    \centering
    \includegraphics[width=0.48\textwidth]{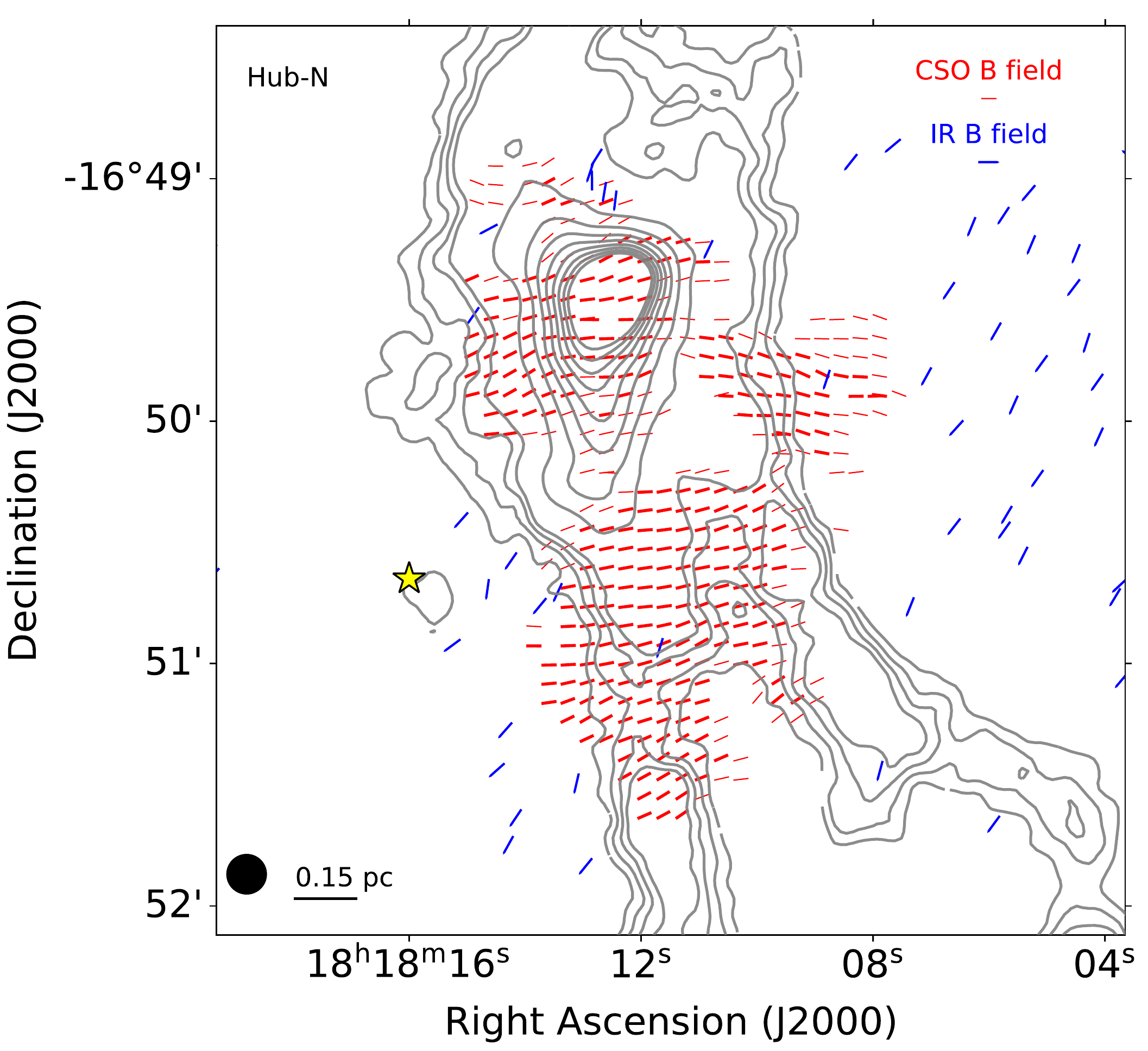}
    \includegraphics[width=0.48\textwidth]{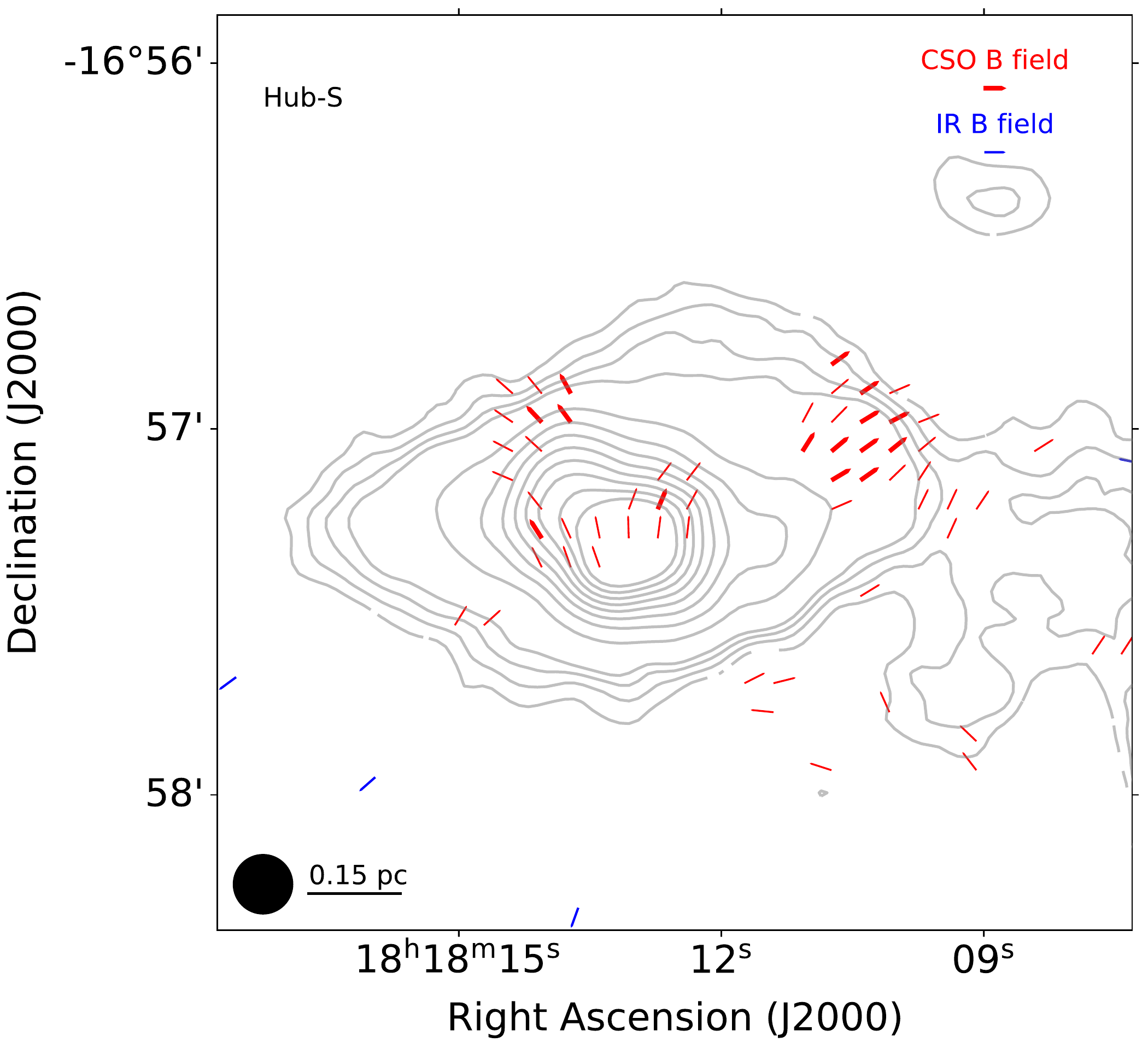}
    \caption{Polarization data toward Hub-N (top panel) and Hub-S (bottom panel) of the IRDC G14.2. Red segments depict the magnetic field (Nyquist sample) observed with the CSO at 350~\mum\ obtained by rotating polarization vectors by 90$\degr$. The width illustrates segments above 3$\sigma_p$ (thick segments) and between 2$\sigma_p$ and 3$\sigma_p$
    (thin segments). Only segments over 15\%/25\% in Hub-N/Hub-S of Stokes I peak intensity are shown. Blue segments depict the magnetic field obtained from near-infrared ($H$-band) observations \citep{Santos2016}. Contours show the CSO 350~\mum dust continuum emission \citep{Busquet2016,Lin2017,Lin2017b}.
 Contour levels are 2, 4, 6, 8, 10, 20, 30, 40, 50, 60, 70, 80, 90 and 100 times the rms noise ($\sim80$~m\jybeam). The CSO beam ($\sim10''$) is shown in the bottom left corner of each panel. The yellow star (top panel) depicts the position of IRAS 18153-1651.
    }
       \label{fig:bfield}
\end{figure}

\begin{figure}[t]
    \includegraphics[width=0.48\textwidth]{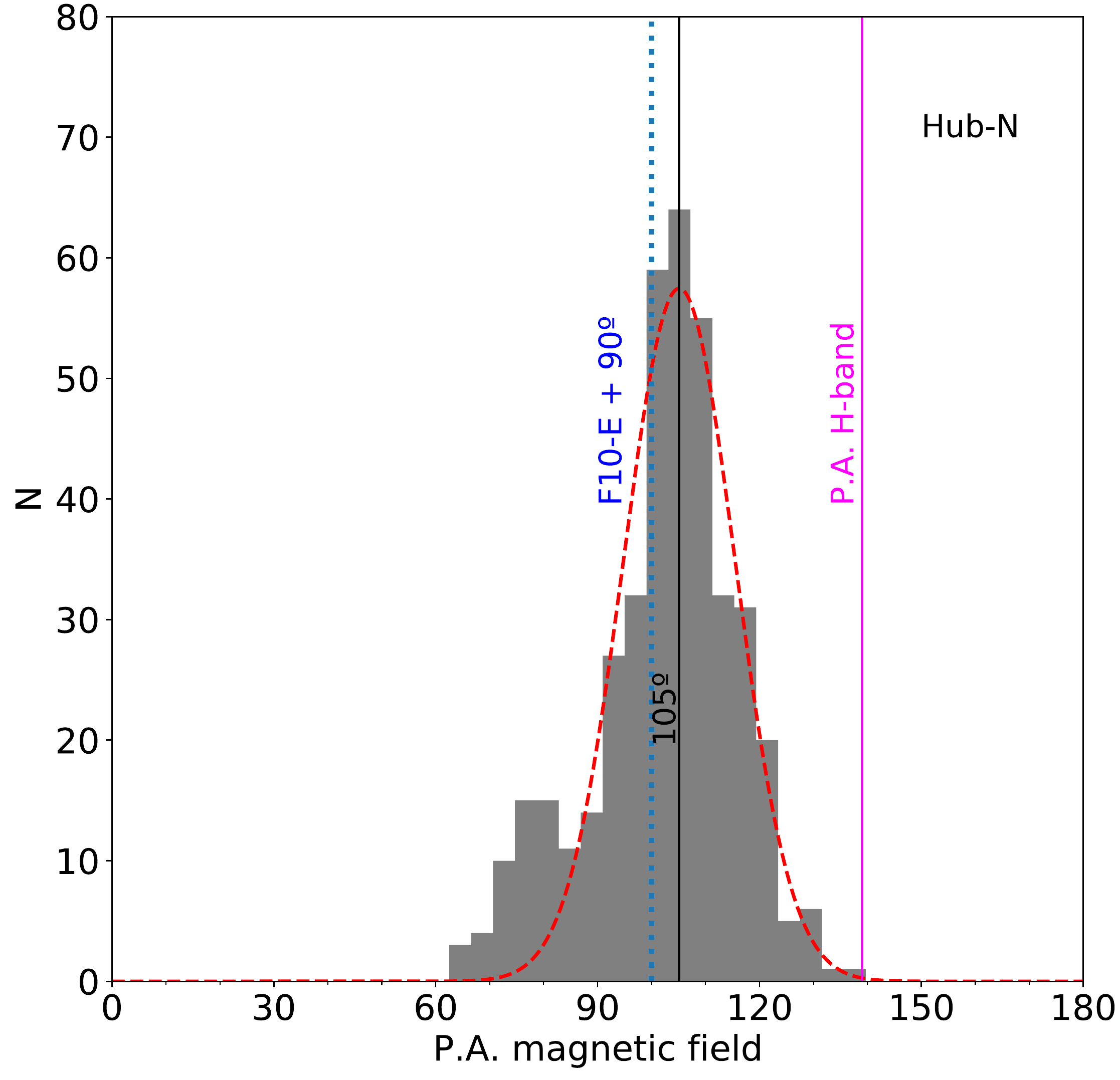}
     \includegraphics[width=0.48\textwidth]{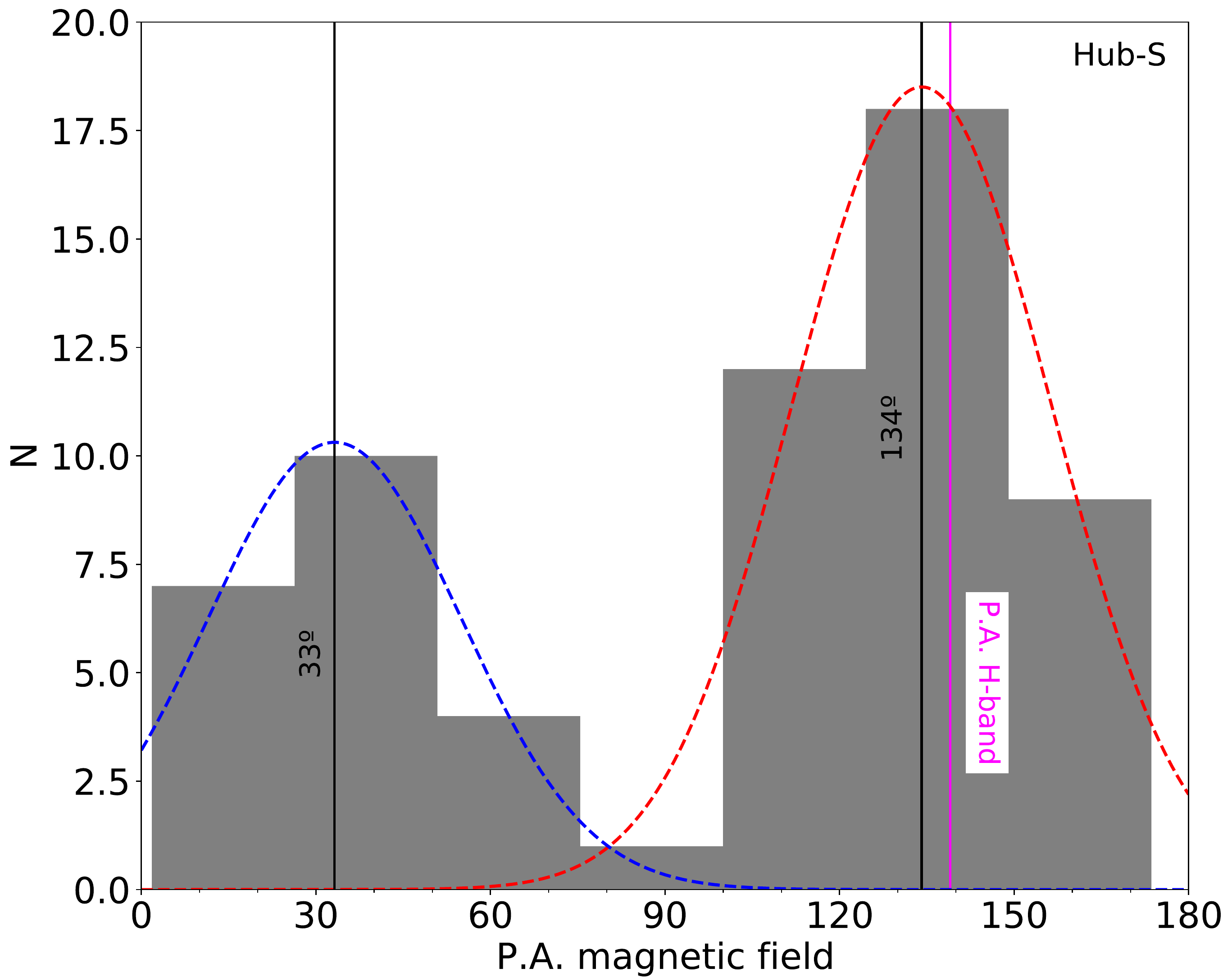}
    \caption{Histogram of the B-field position angles for Hub-N (top) and Hub-S (bottom). 
    Distribution peaks at $105\pm12\degr$ in Hub-N and $33\pm25\degr$/$134\pm26\degr$ in Hub-S, were obtained by a Gaussian fit. The direction perpendicular to the main filament 
   % \LEt{ please remove the following space in "i.e.,".}
    \citep[\ie\ F10-E, $\sim100\degr$,][]{Busquet2013} is indicated by the dotted blue vertical line. The magenta solid line indicates the overall orientation of the polarization in the $H$-band \citep{Santos2016}.}
    \label{fig:pa_histo}
\end{figure}

%%%%%%%%%%%%%%%%%%%%%%%%%%%%%%%%%%%%%%%%%%%%%%%%%%%%%%%%%%%%%%%%%%%%

Figure~\ref{fig:bfield} presents the sky-projected orientation of the magnetic field that is  assumed to be traced by linear polarization.
%\LEt{Please check that I have retained your intended meaning.}. 
Red segments represent the magnetic field direction (polarization segments rotated by 90$^{\circ}$) observed at 350~\mum with an angular resolution of $\sim10''$. Thick segments denote a polarization signal larger than $3\sigma_{\mathrm{P}}$ and thin segments indicate a signal of between $2\sigma_{\mathrm{P}}$ and $3\sigma_{\mathrm{P}}$, where $\sigma_{\mathrm{P}}$ is the noise of the polarization signal. In addition, we only consider segments with a percentage of polarization larger than 0.3$\%$ and with intensity larger than 15\% or 25\% in Hub-N or Hub-S, respectively, of the Stokes I local peak emission. We applied a different intensity threshold because of the different peak intensity in every hub 
%in order 
so as not to remove segments with a S/N of greater than 3-$\sigma$ in hub-N or to keep apparently noisy segments in hub-S. In any case, due to the restrictions applied during the subsequent analysis, which only considers measurements near the peaks, the stricter cut does not affect our analysis.  
The uncertainties of the position angles detected with a S/N of 2-$\sigma$ and 3-$\sigma$ are 17$^{\circ}$ and 10$^{\circ}$ \citep{Naghizadeh1993}. Therefore, we used only values larger than 3-$\sigma$ to compute the average uncertainty in each field. We find that these are 5.8$^{\circ}$ and 7.9$^{\circ}$ for Hub-N and Hub-S, respectively.
In Fig.~\ref{fig:bfield}, blue segments show the magnetic field observed at infrared wavelengths ($H$-band) with the 1.6~m telescope of the Pico dos Dias Observatory \citep{Santos2016}. As shown in Fig.~\ref{fig:bfield} (top panel), the magnetic field structure in Hub-N is rather ordered and uniform, that is, it is mostly along an east--west orientation and roughly perpendicular to the major axis of the hub--filament system direction (\ie\ F10-E following the nomenclature of \citealt{Busquet2013}).  This can be also seen in Fig.~\ref{fig:pa_histo} (top panel), where we present the histogram of the B-field angles. It is clear that there is a prevailing orientation with a dominating single peak 
over a rather small and confined range in B-field angles (mostly between $90$\degr\ and $120$\degr).

We fit a Gaussian to the position angle distribution shown in this figure (the fit is shown as a red line). The peak is found at $105\pm12$\degr. For
comparison, the direction perpendicular to filament F10-E (\ie\ 100\degr) is also depicted in Fig.~\ref{fig:pa_histo}. 
From Fig.~\ref{fig:bfield}, it is clear that a direct point-by-point comparison between the magnetic field orientation inferred from the 350~\mum dust polarization and the $H$-band observations cannot be conducted as there is no good spatial overlap between the two data sets. However, the overall orientation of the polarization in the $H$-band, including all detections over a larger field of view ($\sim40'\times40'$), is $139\pm16\degr$ \citep[see magenta line in Fig.
\ref{fig:pa_histo},][]{Santos2016}, 
which indicates a relatively significant change of the magnetic field direction from cloud to filament scales.

Although the polarization data in Hub-S are scarce (see bottom panel of Fig.~\ref{fig:bfield}), the histogram shown in Fig.~\ref{fig:pa_histo} (bottom panel) presents a bimodal distribution with two peaks: a dominant peak at $134\pm$26\degr\ and a secondary peak around $33\pm$25\degr (values obtained from a Gaussian fit, shown in this panel as a red and blue line). 
Such a bimodal distribution might be reflecting a ``wing-like'' magnetic field structure. There seem to be two wings coming into the main dust continuum peak (which is, presumably, the main gravitational center;  see Fig.~\ref{fig:bfield}-lower panel), with one wing going from the northeast to the southwest and the other wing going from the northwest to the southeast. Despite the large uncertainty on the distribution of polarization angles, the direction of this second (and main) component peaking at $\sim134$\degr\ (west-center region in Fig. \ref{fig:bfield}-bottom panel) is compatible with the main magnetic field orientation seen in Hub-N.

\begin{figure}[t]
    \includegraphics[width=0.5\textwidth]{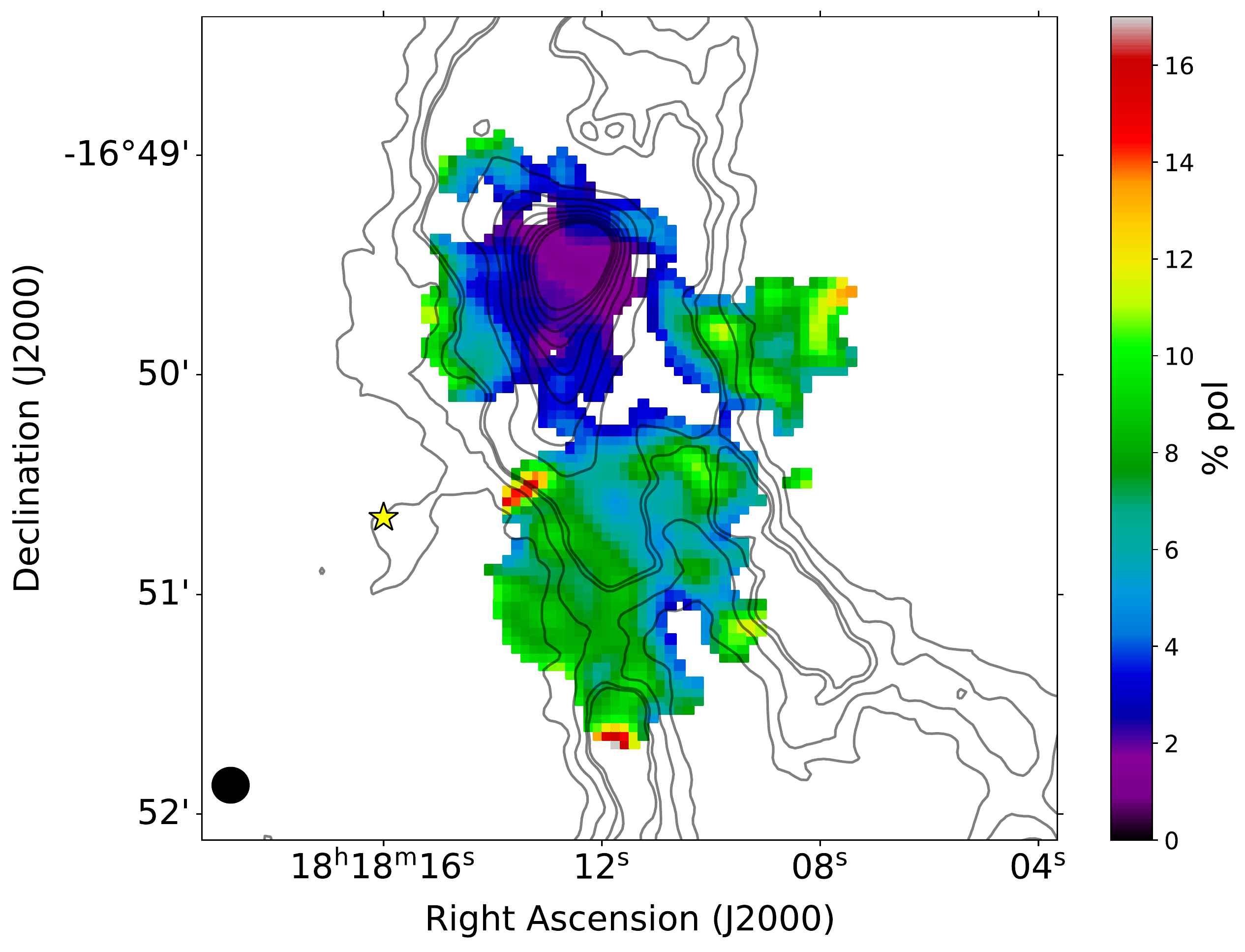}
    \caption{Polarization percentage map using data above 2$\sigma_p$ and lower polarization threshold of 0.3$\%$. Only  segments  over  15\% of Stokes I peak intensity are considered. Contours and markers are the same as in Fig.~\ref{fig:bfield}.
    %\textcolor{blue}{TO DO LIST: Add the position of the IRAS source}
    }
    \label{fig:percentpol_north}
\end{figure}

The polarization percentage map of Hub-N is shown in Fig.~\ref{fig:percentpol_north}. The scarce data detection in Hub-S prevents us from building a similar map toward this region. In Hub-N, the lowest values, namely of $\sim1$\% (see Table \ref{tab:powerlawfit}), are found toward the dust continuum peak. The high-angular-resolution SMA observations of Hub-N reveal a dominant dust continuum core (MM1a) with a mass of $13$~\mo\ and associated with H$_2$O maser activity \citep{Busquet2016}, and hence the core is undergoing gravitational collapse.
Such depolarization may result from a polarization structure that is averaged out by the large CSO beam, similar to the case of the Orion~KL region \citep{Rao1998} and NGC\,1333 IRAS\,4B \citep{Attard2009}. Another very clear case is the successively higher resolution observations in W51, starting from BIMA \citep[3'',][]{Lai2001} to the SMA \citep[0.7'',][]{Tang2009A} and ALMA \citep[0.25'',][]{Koch2018}. Observing the region of depolarization in the BIMA data with the SMA  revealed the collapsing core signature in the B-field structure in W51 e2, and observing the inner SMA depolarization region in W51 e2 with ALMA revealed the likely existence of a pseudo-disk B-field structure.

\begin{figure}[!t]
    \includegraphics[width=0.48\textwidth]{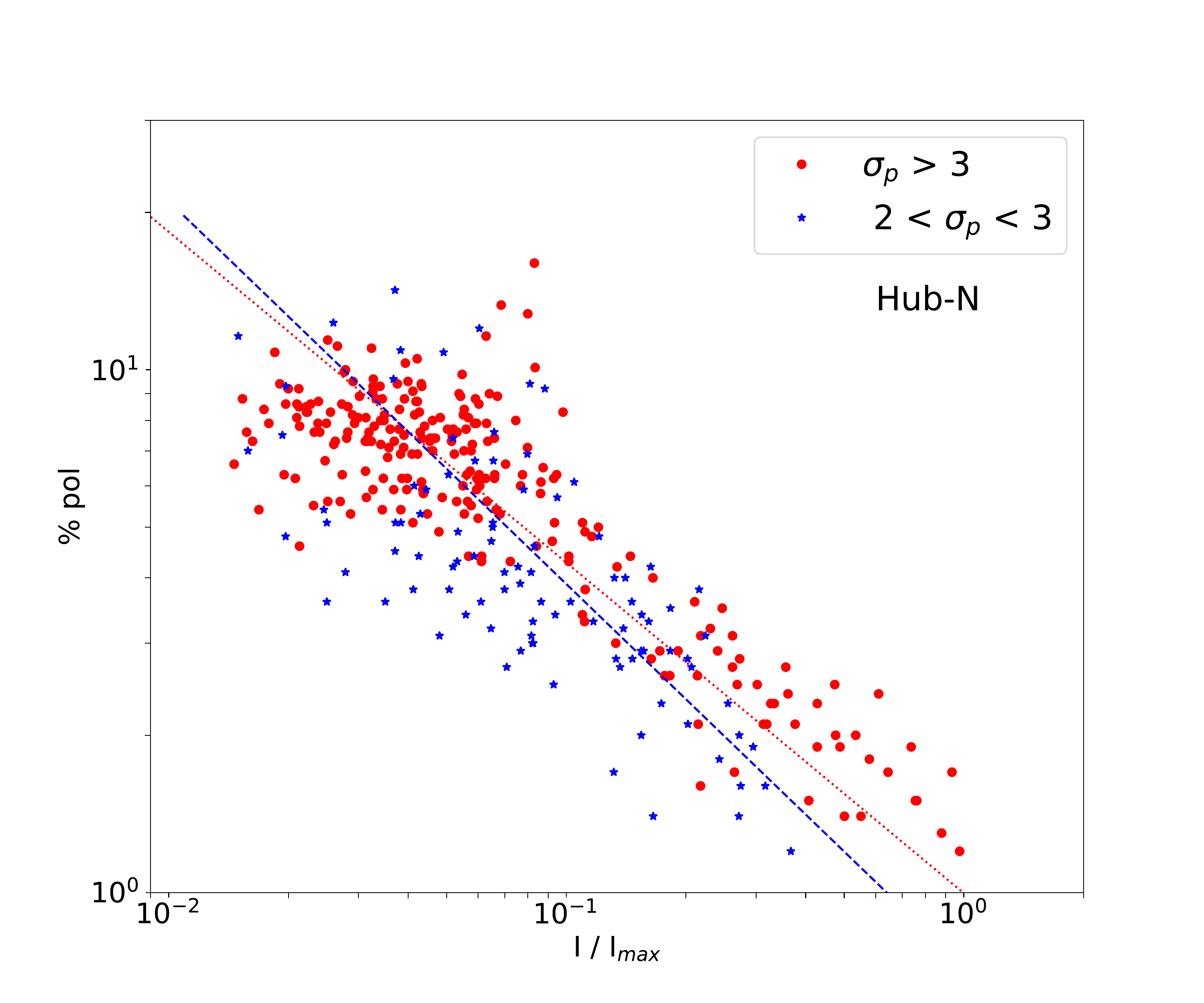}
    \includegraphics[width=0.48\textwidth]{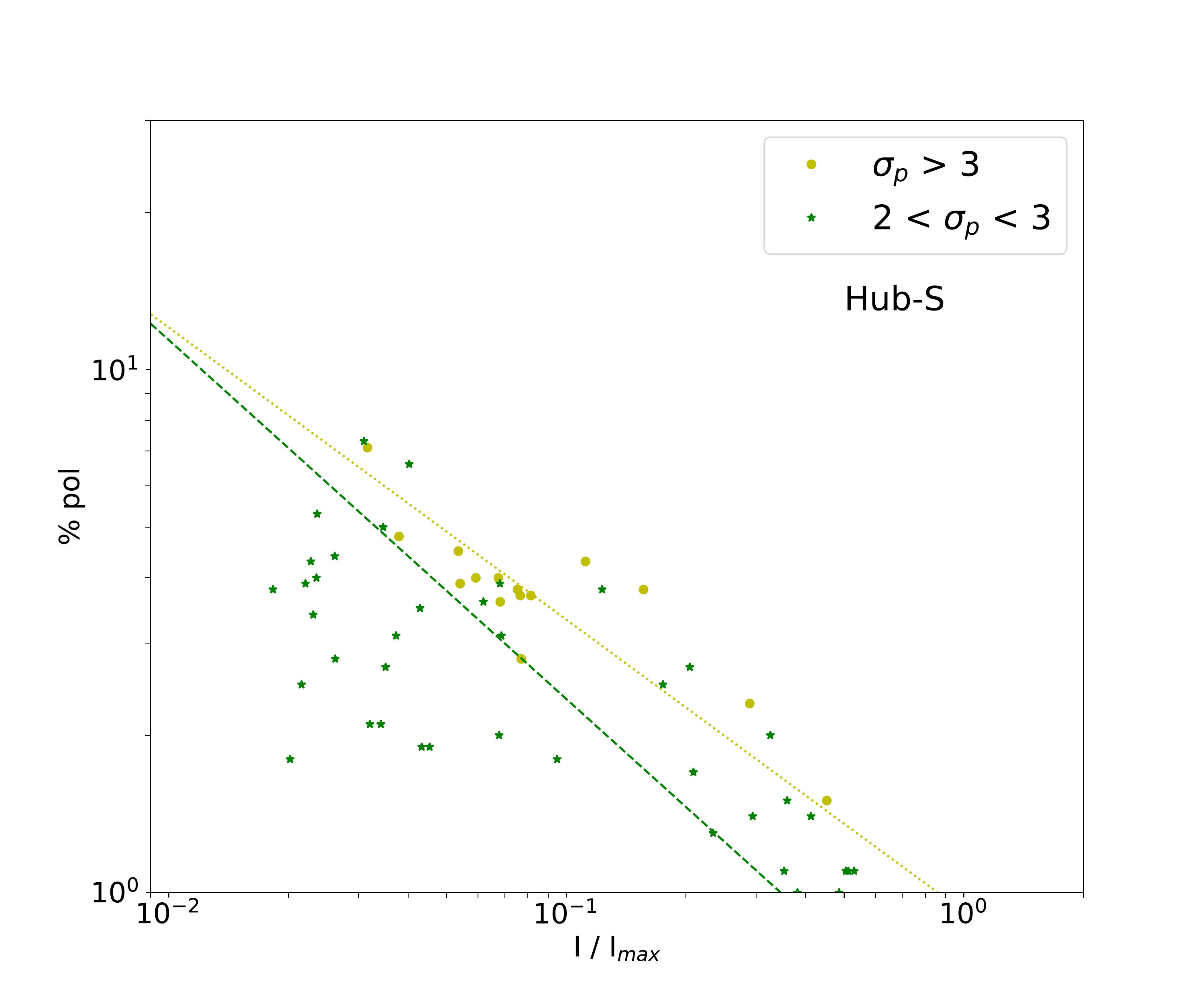}
    \caption{Polarization percentage as a function of the intensity normalized to the peak emission $I/I_{\mathrm{max}}$ for Hub-N (upper panel) and Hub-S (lower panel), computed pixel by pixel. %\LEt{ here for example; please reword avoiding the forward slashes.}
Red and Yellow circles show data $>3\sigma_{\mathrm{P}}$ whereas blue and green stars show data between 2 and $3\sigma_{\mathrm{P}}$ for Hub-N and Hub-S. 
Red and yellow dotted lines and blue and green dashed lines show the power laws that best fit the data above $3\sigma_{\mathrm{P}}$ and the combined data, respectively (see Table~\ref{tab:powerlawfit}). A lower intensity threshold of 3$\sigma$ ($\sigma$=0.08 \jybeam) has been considered.} 
\label{fig:percentpol_south}
\end{figure}

Figure~\ref{fig:percentpol_south} shows the polarization percentage as a function of the intensity normalized to the peak emission, $I$/$I_{\mathrm{max}}$, for both Hub-N (upper panel) and Hub-S (lower panel). 
The polarization percentages in Hub-N and Hub-S extend one order of magnitude, ranging from 1.2~\% and 0.9~\% to 16.0~\% and 7.3~\%, respectively (see Table \ref{tab:powerlawfit}), being slightly lower in Hub-S. Despite the relatively broad scatter of polarization percentages at the low-intensity contours, for both hubs there is an anticorrelation between the polarization percentage and the intensity. 
This anticorrelation can be fitted, without applying any weighting to the values,
 %\textcolor[rgb]{0.984314,0.00784314,0.027451}{can be unweighted fitted\LEt{ the meaning here is unclear; please consider rewording.}} 
with power laws with indices of around $-0.63$ and $-0.56$ for Hub-N and Hub-S, respectively (see Table~\ref{tab:powerlawfit}). In Fig.~\ref{fig:percentpol_south}, we show the power-law fit considering all data points above $2\sigma_{\mathrm{P}}$ (blue and green lines in Fig.~\ref{fig:percentpol_south} for Hub-N and Hub-S) and only data above $3\sigma_{\mathrm{P}}$ (red and yellow lines in Fig.~\ref{fig:percentpol_south} for Hub-N and Hub-S).
%\LEt{ the forward slash is often used to denote a fraction or ratio and so should be avoided for separations such as these; please edit throughout for this change.}.  
Table~\ref{tab:powerlawfit} reports the power-law indices considering these two data sets.
Similar slopes have been found in the filamentary IRDC G34.43$+$00.24 \citep{Tang2019} and at core scale in high-mass star-forming regions  \citep[\eg][]{Koch2018}.

This dependence of the polarization fraction on the dust intensity is typically used to infer dust grain alignment efficiency in star-forming regions \citep[see \eg][]{Whittet2008,Pattle2019}. However, we stress that the observed depolarization toward the CSO dust peak in Hub-N may be due to beam-smearing effects, which prevents us from drawing any firm conclusions on dust alignment efficiency.

%%%%%%%%%%%%%%%%%%%%%%%%%%%%%%%%%%%%%FIGURES%%%%%%%%%%%%%%%%%%%%%%%%%%%%%%%%%%%%%%%%%%%%%%%%%%%%%%%

\begin{table}[!t]
    \centering
    \caption{Polarization percentage properties} 
    \label{tab:powerlawfit}
    \begin{threeparttable}[]
    \begin{tabular}{c c c c c c c}
    \hline\hline
        Field & Threshold \tnote{a} & Median  & Max & Min & Std \tnote{b} & Slope \tnote{c} \\
        & ($\sigma_p$) & ($\%$) & ($\%$) & ($\%$) & ($\%$) & \\
        \hline
         Hub-N & 3 & 7.2 & 16.0 & 1.2 & 2.5 &$-$0.63\\
         Hub-N & 2 & 4.8 & 14.2 & 1.2~ & 2.8 &$-$0.73\\
         Hub-S & 3 & 3.8 & \phn 7.1 & 1.5 & 1.2 & $-$0.56\\
         Hub-S & 2 & 2.7 & \phn 7.3  & 0.9  & 1.6  & $-$0.68\\
    \hline
    \end{tabular}
    \begin{tablenotes}
    \item[a] We only consider vectors with intensity greater than this threshold for the calculation of statistical values.
   % Threshold from which segments are considered for statistic\LEt{ the intended meaning here is unclear; please consider expanding slightly.}.
    \item[b] Standard deviation of the polarization percentage distribution.
    \item[c] Power-law index for fitting between normalized intensity and polarized emission percentage (see Fig. \ref{fig:percentpol_south}).
    \end{tablenotes}
     \end{threeparttable}
\end{table}

%%%%%%%%%%%%%%%%%%% ANALYSIS %%%%%%%%%%%%%%%%%%%%%%%
\section{Analysis}\label{sec:ana}

\subsection{Approach and analysis techniques}\label{sec:methods}

In this work, we applied the method developed by \cite{Koch2012, Koch2012b} to estimate the local magnetic field-to-gravity force ratio $\Sigma_{\mathrm{B}}$. 
This method is motivated by the relationship between the magnetic field and the intensity gradient, and it begins with the force equation in the ideal magnetohydrodynamics (MHD) environment:

\begin{equation}
\rho \left( \frac{\partial}{\partial t}  + \Vec{v} \cdot \nabla \right)\Vec{v}
= - \nabla \left(P + \frac{B^2}{8\pi}\right)
-\rho \nabla \phi + \frac{1}{4\pi} \left(\Vec{B}\cdot\nabla\right) \Vec{B}, \label{eq:MHD}
\end{equation}

\begin{figure*}[!ht]
    \centering
    \includegraphics[width=\textwidth]{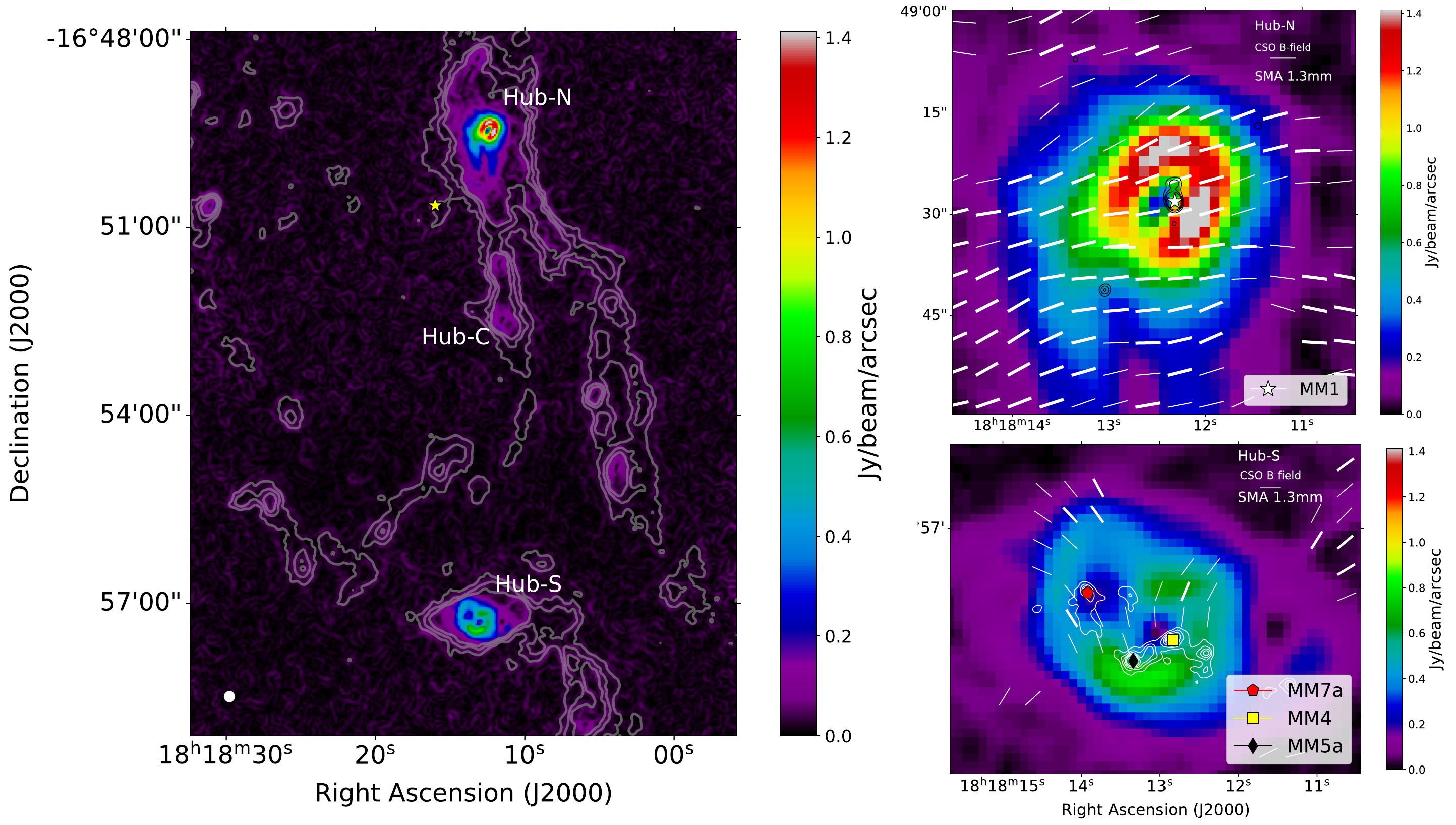}
    \caption{\textit{Left:} CSO intensity gradient magnitudes at 350~\mum in units of \jybeam\,arcsec$^{-1}$ (color scale) overlaid on the CSO dust continuum emission at 350~\mum (contours, \citet{Busquet2016,Lin2017}). Contour levels are 3, 6, 9, and 12 times the rms noise of 80~m\jybeam. The yellow star depicts the position of IRAS\,18153$-$1651. 
    The white circle in the bottom left corner depicts the CSO beam size ($\sim10''$). Hub-N, Hub-S, and the Hub-C candidate identified by \citet{Chen2019} are labeled.
    \textit{Right:} Close-up images of the intensity gradient toward Hub-N (top panel) and Hub-S (bottom panel) overlaid on the SMA 1.3~mm dust continuum emission (black and white contours) at an angular resolution of $\sim1\farcs5$ \citep{Busquet2016}. Contour levels range from $12\sigma$ to $30\sigma$ in steps of $6\sigma$, and from $30\sigma$ to $60\sigma$ in steps of $15\sigma$, in Hub-N and similar but starting from $6\sigma$ in Hub-S, where $\sigma$ is the rms of the map, $\sim1$~m\jybeam.
    White segments show the CSO B-field orientation, similar to Fig.~\ref{fig:bfield}. 
    White star (right upper panel), 
    red pentagon, black diamond, and yellow square (right bottom panel) depict MM1a, MM7a, MM4, and MM5a cores, respectively. 
}
    \label{fig:fulgrad}
\end{figure*}

\noindent
where $\rho$ is the dust density, $\Vec{v}$ is the dust velocity, \Vec{B} is the magnetic field, $P$ is the hydrostatic dust pressure, $\phi$ is the gravitational potential resulting from the total mass contained in the star-forming region, and $\nabla$ denotes the gradient. As shown by \cite{Koch2012}, we can transform Equation~\ref{eq:MHD} into:

\begin{equation}
    \rho v \frac{\partial v}{\partial s_v}\Vec{e_{s_v}} +
    \rho v^2 \frac{\partial \Vec{e_{s_v}}}{\partial s_v} =
    - \frac{\partial P}{\partial s_p} \Vec{e_{s_p}}-
    \rho \frac{\partial \phi}{\partial s_{\phi}} \Vec{e_{s_{\phi}}} +
    \frac{1}{4 \pi} B^2 \frac{1}{R} \Vec{n}, \label{eq:transformed}
\end{equation}

\noindent
where generalized coordinates $s_v$, $s_P$, $s_{\phi}$, $s_B$
along the directions of the unity vectors $\Vec{e_{s_v}}$, $\Vec{e_{s_P}}$, $\Vec{e_{s_{\phi}}}$, and $\Vec{e_{s_B}}$ are used,  
and the unity vector $\Vec{n}$ is directed normal to a magnetic field line that is along $\Vec{e_{s_B}}$. Here,$1/R$ is the curvature of the magnetic field line.

The interaction between hydrostatic pressure, gravitational potential, magnetic field force, and the resulting motion are described by Equation~\ref{eq:transformed}. In \cite{Koch2012}, the authors assume that 
the observed distribution of emission intensity is attributable to the  transport of matter driven by a combination of the forces stated above. In other words, a change in emission distribution is the result of all combined forces. With this assumption, an emission intensity gradient measures the resulting direction of motion, expressed by the inertial
term on the left-hand side of equations (\ref{eq:MHD}) and (\ref{eq:transformed}). With this, and
identifying the various force terms in 
Equation~\ref{eq:transformed} in an observed map in two dimensions \citep[see Fig.~3 in][]{Koch2012}, and solving for the magnetic field, we can derive the following equation:

\begin{equation}
    B = \sqrt{\frac{\sin \psi}{\sin \alpha} \left(\nabla P + \rho \nabla \phi \right) 4\pi R}, \label{eq:bfield}
,\end{equation}

\noindent
where $\psi$ and $\alpha$ are the angles in the plane of the sky between the orientation of local 
gravity and intensity gradient, and those between polarization and  intensity gradient, respectively.

In addition, making use of the magnetic field tension force, and the gravitational and pressure forces:
\begin{equation}
\begin{array}{cl}
     F_B & = B^2 / 4 \pi R  \\
     | F_G + F_P | & = | \rho \nabla \phi + \nabla P |. 
\end{array}
\end{equation}

Equation~\ref{eq:bfield} can be written as 
\begin{equation} \label{eq:sigmaB}
    \left(\frac{\sin \psi}{\sin \alpha} \right) =
    \left(\frac{F_B}{|F_G + F_P|} \right) \equiv 
    \Sigma_B ,
\end{equation}

\noindent
where $\Sigma_B$ is a concept introduced by \cite{Koch2012} to define the 
local magnetic field significance in the presence of gravity and any hydrostatic pressure. 
$\Sigma_B$ evaluates the relative importance between the magnetic field tension force and the other forces involved. In star-forming regions, the thermal pressure term $\nabla P$ is typically small and negligible as compared to gravity. Therefore, $\Sigma_B$ provides a criterion to distinguish a scenario where the magnetic field can prevent gravitational collapse and infall ($\Sigma_B$ > 1) or not ($\Sigma_B$ < 1).
We note that all of the above equations can be evaluated locally at any position in a map where a magnetic field orientation is detected and where a local direction of gravity and an intensity gradient can be calculated based on a detected (dust) emission. These equations therefore allow us to construct maps of the magnetic field significance
$\Sigma_B$ and field strength $B$, and they provide a way of assessing where star formation is likely to  preferentially occur. %\LEt{Please check that I have retained your intended meaning.}

The angle $|\delta|$ \footnote{For the purpose of the analysis here, it is sufficient to consider | delta | $\leq$ 90 deg. 
We note that the angle delta can additionally be given a sense of orientation,
-90 deg $\leq$ delta $\leq$ 90 deg, 
as explained in \cite{Koch2013}. This additionally captures systematic differences
in the deviation of the magnetic field clock-wise or counterclock-wise from an 
intensity gradient, as e.g., in an hour-glass morphology that reveals characteristic      
positive and negative deviations.} \citep[$=\pi$/2 - $\alpha$; see Fig.3 in][]{Koch2012} is the angle between a projected magnetic field orientation and an intensity gradient direction. \cite{Koch2012} propose the angle $|\delta|$ as a diagnostic to measure the dynamical role of the magnetic field. In \cite{Koch2013}, the authors develop the physical meaning of the angle $|\delta|$ by exploring its behavior at different scales
across a large sample of about 30 sources observed with the SMA and the CSO. Further statistical evidence based on a sample of 50 sources is given in \cite{Koch2014}.
These latter authors conclude that a $|\delta|$--map always allows for an approximation of the more refined $\Sigma_B$ map. Given the insufficient polarization data for Hub-S, except for its intensity gradient map, in the following sections we present the results obtained for Hub-N.

%%%%%%%%%%%%%%%%%%%%%% RESULT OF THE ANALYSIS %%%%%%%%%%%
\subsection{Polarization versus intensity gradient}

We used the 350~\mum dust continuum map presented in \citet{Busquet2016} and \citet{Lin2017} to build an intensity gradient map. First we computed the numeric spatial derivative \footnote{We used the Python tool numpy.gradient to compute the gradient map.}, from which we got \textit{X} and \textit{Y} vector components for every pixel. We then computed the vector magnitude encompassing the whole IRDC complex G14.2 (see Fig.~\ref{fig:fulgrad}). The highest values of the intensity gradient magnitudes are found toward the hubs. Obviously, at the peak of the dust emission, the intensity gradient magnitude decreases, which results in a ring-like morphology with maximum intensity gradient around the inner hole. As can be seen in Fig.~\ref{fig:fulgrad}, we can distinguish the filamentary structures detected in NH$_3$ by \citet{Busquet2013}. Along the filaments, we find various regions where the intensity gradient slightly increases, coinciding with the dust clumps 
and the Hub-C candidate 
identified in N$_2$H$^+$ \citep{Chen2019}.

\begin{figure}
    \centering
 \includegraphics[width=0.48\textwidth]{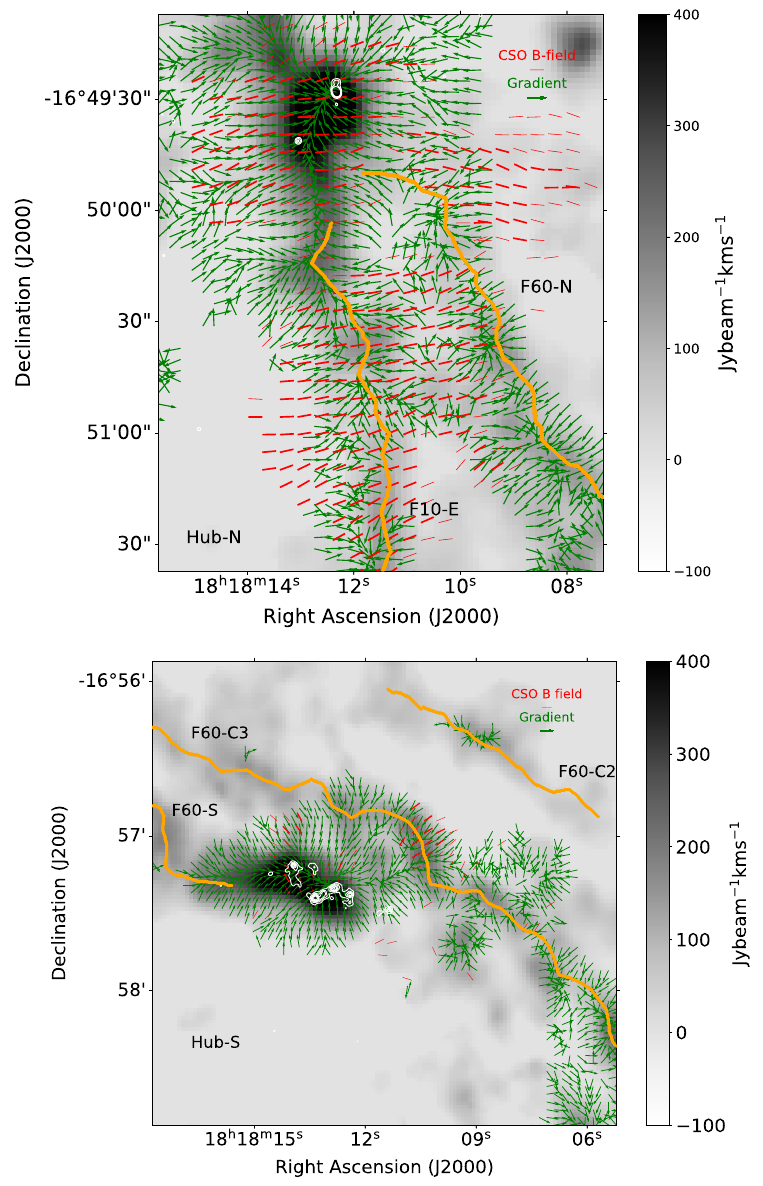}
    \caption{NH$_3$ moment-zero map (grayscale; \citealt{Busquet2013}) overlaid on the magnetic field segments (red) and the intensity gradient vectors (green, uniform length) from dust continuum for Hub-N
    (upper panel) and Hub-S (lower panel). 
    We have restricted the representation of the gradient to the areas where the emission exceeds three times the rms noise of the continuum map. 
    White contours correspond to the SMA 1.3~mm dust continuum emission \citep{Busquet2016},
    similar to Fig.~\ref{fig:fulgrad}.
    The orange lines depict the spine of the filaments 
    identified in N$_2$H$^+$ by \citet{Chen2019} and are labeled according to \citet{Busquet2013}.
    }
    \label{fig:fil-spine}
\end{figure}

\begin{figure*}[!ht]
    \includegraphics[width=0.98\textwidth]{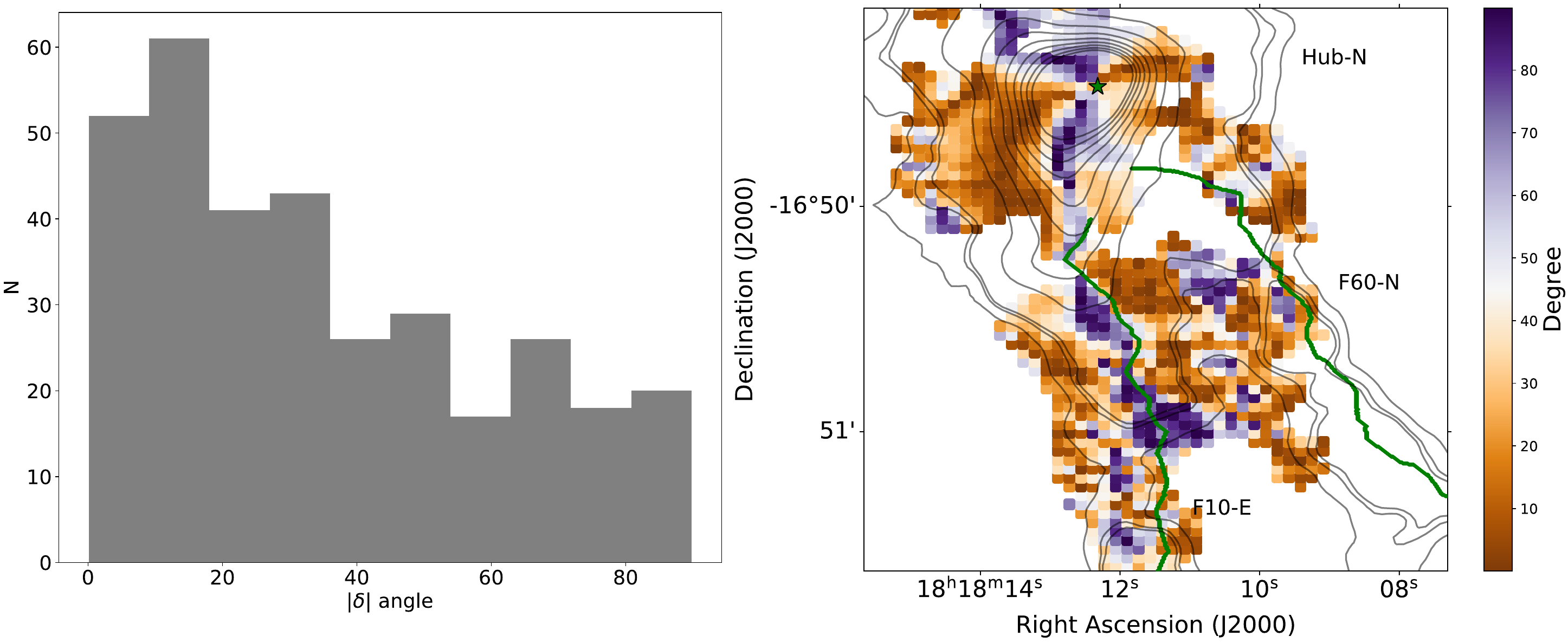}
    \caption{ 
    \textit{Left:} Histogram of the absolute difference, $\lvert\delta\rvert$, between the position angles
   % \LEt{ if wishing to introduce this acronym, this must be done in the main text upon first use and then it must used exclusively thereafter.}  
   of magnetic field and 
    intensity gradient in the range from $0\degr$ to $90\degr$. 
    \textit{Right:} $|\delta|$ map toward Hub-N (color scale) overlaid on the 350~\mum dust emission. Contour levels range from 2$\sigma$ to 10$\sigma$ in steps of $2\sigma$ and from 10$\sigma$ to $100\sigma$ in steps of 10$\sigma$, where $\sigma$ is the rms noise of the map ($\sim80$~m\jybeam). The green star and the green solid line depict the MM1a core position and the spine of the filament, respectively, similar to Fig.~\ref{fig:fil-spine}. }
    \label{fig:deltamap_North}
\end{figure*}

The right panels in Fig.~\ref{fig:fulgrad} show the close-up images of Hub-N (top panel) and Hub-S (bottom panel) overlaid with the SMA 1.3~mm dust continuum emission from \citet{Busquet2016}. In addition, the magnetic field distribution observed with the CSO at 350~\mum has been included. The high-angular-resolution ($\sim1\farcs5$) SMA 1.3~mm dust continuum images reveal that the two hubs present different levels of fragmentation. While Hub-N contains 4 mm condensations, Hub-S displays a higher fragmentation level, with 13 mm condensations within a field of view of 0.3~pc \citep[][see also Table~\ref{tab:polresults}]{Busquet2016}. Moreover, the spatial distribution of the 1.3~mm dust emission is significantly different in the two hubs. Hub-N presents a bright, centrally peaked dust condensation (MM1a according to the nomenclature of \citealt{Busquet2016}) with only a few and much fainter dust cores in its close vicinity \citep{Busquet2016,Ohashi2016}. 
On the other hand, as shown by \citet{Busquet2016} and \citet{Ohashi2016}, the dust cores in Hub-S are more spread apart. They appear to be clustered in two different spatial locations, a snake-shape structure coinciding with the peak position of the CSO 350~\mum emission, and the second group of dust cores located toward the northeast following the elongation seen in the CSO 350~\mum dust continuum map. This behavior is clearly reflected in the close-up images of the intensity gradient (see Fig.~\ref{fig:fulgrad}, right panels). While in Hub-N the intensity gradient displays a single local minimum coinciding with the bright dust continuum core MM1a, in Hub-S there are two clear minima, matching the spatial distribution at small scales. Interestingly, these close-up images reveal that the prevalent east--west magnetic field direction in Hub-N slightly deviates when approaching the bright dust continuum core MM1a. 
\cite{Tang2019} find this kind of small deviation in G34 in the MM1 core, which is also the core with the most dominant magnetic field; it shows no fragmentation compared to the other cores in G34. 
Regarding Hub-S, Fig.~\ref{fig:fulgrad}   (bottom-right panel) shows that the two components of the bimodal distribution in B-field angles (see Fig.~\ref{fig:pa_histo}-bottom panel) resemble the two incoming wings which appear to converge towards the two local minima in the map of the magnitude of the intensity gradient.
Specifically, the northeast component points toward MM7a, which is the most massive core ($\sim10$~\mo) of the northeast group, whereas the second component (coming from the northwest and southeast) points toward the snake-like structure, in which MM4 ($M\simeq15$~\mo) and MM5a ($M\simeq18$~\mo) are the most massive cores \citep{Busquet2016}. These results suggest a scenario in which the magnetic field in each hub is dragged by the collapsing cores, while at the same time the incoming wing-like and organized B-field morphology might be hinting at accretion zones guided by the B-field.

We note that the intensity gradient can be a subtle tracer of changes in emission, and can therefore localize weak (not yet resolved) maxima and minima in coarser emission.
These local minima (such as in the case of Hub-N and Hub-S) can be signposts for fragmentation structures that  can be detected and resolved with higher-resolution observations.

In order to see whether or not we are looking at the same depth in the two hubs, we estimated the optical depth at 350~\mum. Adopting the physical parameters inferred by \citet{Busquet2016} (see their Table~5), we obtained similar values for optical depth ($\tau_{\mathrm{350\mu m}}\simeq1.6$) in both hubs. Thus, we are confident that differences in the dust polarization are not due to different optical depths.

Figure~\ref{fig:fil-spine} shows the dense gas emission traced by the NH$_3$ molecule toward the two hub-filament systems in G14 from \citet{Busquet2013}, highlighting the extracted filament spine (orange line) identified through N$_2$H$^+$ emission by \citet{Chen2019}. We additionally overlaid the magnetic field segments (in red) and the intensity gradient vectors (in green). 
At large scales, and both east and west of Hub-N and its two associated filaments,
the magnetic field is roughly perpendicular to the major axis of these dense structures, approximately following the direction of the intensity gradient. Close to the dense structures, the magnetic field appears to still be perpendicular to the direction of the  filaments with only a small variation close to the peak emission of Hub-N. In contrast, close to the filaments (especially toward F10-E), the intensity gradient appears to bend, becoming almost parallel to the filament spine, with the intensity gradient pointing toward Hub-N, that is, toward the gravitational potential (see Fig.~\ref{fig:fil-spine}, top panel). In fact, locally converging intensity gradient directions also appear to trace the spine of the extracted filaments, that is, in many places the  intensity gradients converge from left and right towards a central line which overlaps or is near an extracted filament.  We note that, because an intensity gradient map naturally identifies local and global minima and maxima, it will reveal structures similar to any filament-extraction routine that is searching for embedded denser structures.
The bottom panel of Fig.~\ref{fig:fil-spine} shows the same plot for Hub-S. At large distances of Hub-S, the magnetic field is perpendicular to filament F60-C3, following mostly the same direction as the intensity gradient. 
At some point, and when approaching the hub (\ie\ the gravitational potential) the intensity gradient in both the eastern and western sides of Hub-S converge toward it. More specifically, the intensity gradient on the eastern side is perpendicular to F60-S and becomes parallel to the spine when entering the dense gas close to Hub-S where it penetrates
%\LEt{Please check that I have retained your intended meaning.} 
the hub up to the group of dust cores located toward the northeast. 
We observe the same behavior toward the west ($\sim$ 18h18m12s -16º57'20''), with the intensity gradients from the northwest and from the southwest both pointing toward the snake-like structure.
In this case, the magnetic field seems to follow the behavior of the gradient.

Figure~\ref{fig:deltamap_North} presents the results obtained from the analysis of the angle $|\delta|$. 
A significant number of values of $|\delta|$ are below 37$^\circ$ (about 60\%), as shown in the histogram in the left panel, whereas the rest of the values are distributed roughly uniformly above this value.
This is a consequence of the fact that around the main filament the intensity gradient follows the magnetic field lines leading to an elongated structure. This behavior is clearly seen in the pixel-by-pixel map for the angle $|\delta|$ shown in Fig.~\ref{fig:deltamap_North} (right panel). As can be seen in this figure, $|\delta|$ appears small in the outer zones, both from the east and from the west, of the central north--south ridge. 
This might indicate that the magnetic field is channeling material towards the central ridge from both sides and might also indicate that the B-field is initially dragged by the infalling motion and aligned with it.
The rather uniform magnetic field together with small $|\delta|$ values also suggests that the magnetic field, the gravitational pull and possibly also large converging flows toward the ridge are all %\LEt{ without the parentheses this "all" would make sense.}
mostly aligned. It should be noted that larger deviations between the intensity gradient and the magnetic field, that is, larger values of $|\delta|$ occur along the inner spine of the north--south ridge. 
These are locations where gravity has not yet (fully) overwhelmed and forced the alignment of the magnetic field. 

Figure \ref{fig:sigmaB_North} shows the $\Sigma_B$--map for the Hub-N, which represents the relative significance of the magnetic field tension force against gravity, following equation \ref{eq:sigmaB}, where $\Sigma_{B}$ > 1 or $\Sigma_{B}$ < 1 indicate that the magnetic field can or cannot prevent the gravitational collapse, respectively.
%\LEt{Please check that I have retained your intended meaning.}.
According to Fig.~\ref{fig:sigmaB_North}, $\Sigma_{B}<1$ both east and west of the CSO 350~\mum dust emission peak, and also throughout 
most of the inner hub area, and hence gravity dominates over the magnetic field force. 
Values of $\Sigma_{B}\simeq1$ (and slightly larger) are found along a very confined structure, that is, along the north--south ridge, 
where $\Sigma_{B} > 1$ in several positions.
We note the clear spatial correlation between $|\delta|$ and $\Sigma_B$-maps. Large values in $|\delta|$ generally also mean large values in $\Sigma_B$, and low values in $|\delta|$ are associated with low values in $\Sigma_B$ . This fact confirms the angle $|\delta|$ as a diagnostic to measure the dynamical role of the magnetic field.

In summary, for Hub-N the analysis of the intensity gradient together with the $|\delta|$ and $\Sigma_B$ maps indicate that, in general, gravity dominates the magnetic field. The intensity gradient is parallel to the magnetic field lines, both east and west of the dust emission peak, with the magnetic field being perpendicular to the major axis of the hub-filament system, suggesting that magnetic field is channeling material toward the central ridge or that the B-field is being dragged by gravity. On the other hand, this ridge appears to be dominated by the magnetic field ($\Sigma_B>1$). When approaching the dust continuum peak, such uniform distribution is somewhat perturbed, with the magnetic field being roughly perpendicular to the intensity gradient. 
In Hub-S, although the computation of the $|\delta|$ and $\Sigma_{B}$ maps
is less viable because of the more disconnected and isolated polarization detections, we clearly see
that the magnetic field points toward each of the two minima identified in the intensity gradient map, which coincides with the most massive dust cores identified at high angular resolution. This scenario is compatible with the interpretation that the gravitational collapse is dragging the magnetic field toward the collapsing cores.

%%%%%%%%%%%%%%%%%%%%%MAGNETIC FIELD STRENGTH%%%%%%%%%%%%%%%%%%%%

\subsection{Magnetic field strength}
\label{s:CFmethod}

In a previous work, \citet{Busquet2016} investigated the interplay between the different agents acting on the fragmentation process in the two hubs of the IRDC G14.2. These latter authors investigated the underlying density and temperature structure of these hubs and find remarkably similar physical properties. Additionally, all the inferred physical parameters, such as the level of turbulence and the magnetic field strength, are notably similar as well. However, it is important to point out that the magnetic field strength in \citet{Busquet2016} was obtained by scaling the magnetic field inferred from near-infrared $H-$band observations of interstellar polarization of background starlight \citep{Santos2016}, which traces the magnetic field in the diffuse gas surrounding the filaments and hubs of the IRDC complex. The CSO 350~\mum dust polarization data provide the magnetic field distribution penetrating into the dense infrared-dark area, and therefore provide a more accurate estimation of the magnetic field strength deep into the hubs.
It is important to mention that the magnetic field strength in Hub-N was estimated using a 3$\sigma_p$ threshold whereas a threshold of 2$\sigma_p$ was adopted for Hub-S as it contained fewer polarimetric detections. 
Furthermore, all calculations related to the magnetic field strength and derived magnitudes are restricted to an area of $0.15$~pc in radius (centered on the phase center of the SMA maps) for consistency with \cite{Busquet2016}.

\subsubsection{Method~I: Davis-Chandrasekhar–Fermi (DCF) \label{sec:methodI}}

The magnetic field strength in the plane of the sky ($B_{\mathrm{pos}}$) can be estimated from the DCF technique \citep{DavisPhysRev,ChandrasekharFermi1953} using Equation~2 of \citet{Crutcher2004}:

\begin{equation}
    B_{\text{pos}} = Q \sqrt{4\pi \rho} \frac{\sigma_{\mathrm{1D,nth}}}{\sigma_{\mathrm{P.A.}}}\approx 9.3 \frac{\sqrt{n(\text{H}_2}) \Delta\,v}{\langle \sigma_{\mathrm{P.A.}}\rangle}~\mu G
,\end{equation}

\noindent
where $\rho$ is the gas density, $\sigma_{\mathrm{1D,nth}}$ is the turbulent velocity dispersion, $\sigma_{\mathrm{P.A.}}$ is the dispersion in polarization position angles (in degrees), $n$(H$_2$) is the molecular hydrogen number density (in cm$^{-3}$), and 
$\Delta\,v = \sigma_{\mathrm{1D,nth}}\,\sqrt{8\,\mathrm{ln}(2)}$
(in \kms) is the FWHM line width. 
We included the numerical correction factor $Q=0.5$ \citep{Ostriker2001}, which yields a good estimation of the $B_{\mathrm{pos}}$ strength when the dispersion in polarization angles is $<25\degr$.

Table~\ref{tab:polresults} reports the relevant physical parameters inferred in \citet{Busquet2016} that have been used to obtain the magnetic field strength, the mass-to-flux ratio, and the Alfvén Mach number (see Section~\ref{sec:masstoflux}).

In Table \ref{tab:polresults} we list the magnetic field strength computed using the DCF technique for Hub-N and Hub-S, finding 0.8 and 0.2~mG, respectively. 
The uncertainties on $B_{\mathrm{pos}}$ were estimated by propagating the errors in $\Delta\,v$, $n$(H$_{\mathrm{2}}$), and $\sigma_{\mathrm{P.A.}}$
resulting in an error of a factor of $\sim$2/1.5, that is $\pm$ 0.8 and 0.1~mG for Hub-N and Hub-S, respectively. In Table~\ref{tab:polresults}, we also show the magnetic field angular dispersion for both Hub-N ($\sim7.8\degr$) and Hub-S ($\sim29\degr$) computed as the standard deviation. Due to large dispersion in Hub-S (larger than 25\degr), we can consider this result as an upper limit given the numerical correction factor assumed ($Q$=0.5).
In addition, the further analysis of Hub-S is not yet conclusive with the present polarization data because of their rather disconnected and isolated detection.

We included the modified DCF technique developed by \cite{Heitsch2001}, 
which  is no longer dependent on low angle dispersion because it does not use the tangent angle approximation. Table~\ref{tab:polresults} summarizes the estimation of the B-field strength in Hub-N, which reaches a slightly lower value than the previous approximation, $B_{\mathrm{pos}}\simeq0.7$~mG. Regarding Hub-S, we find approximately the same value as with the DCF technique (0.1~mG).

\begin{figure}[ht]
    \includegraphics[width=0.48\textwidth]{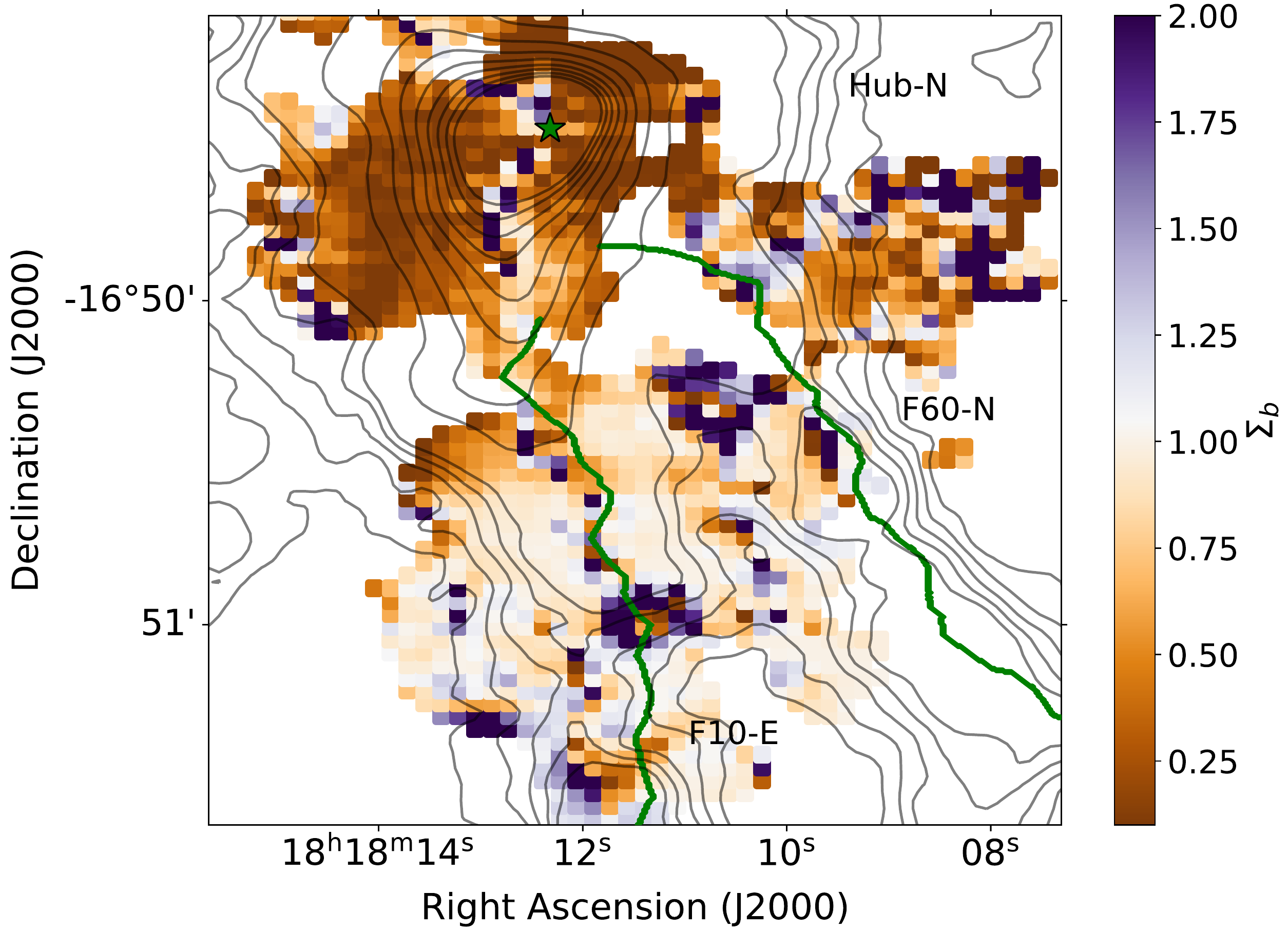}
    \caption{Spatial distribution of the B-field-to-gravity force ratio, $\Sigma_{B}$, toward Hub-N.
    Contour levels, symbols, and lines are the same as in Fig.~\ref{fig:deltamap_North}. 
    We note that while in the hub there is a smooth transition of the values of $\Sigma_B$, in the boundaries of the hub and in the south of the map there are pixels with outliers. In these regions, there is a noisy intensity gradient (see Fig. 6) that is likely due to the absence of a dominant
    center of gravity at these positions.
    }
    \label{fig:sigmaB_North}
\end{figure}

\subsubsection{Method~II: polarization--intensity gradient technique}

According to the method illustrated in Section~\ref{sec:methods}, we proceed to estimate the magnetic field strength based on the relative orientation between local magnetic field and the estimated local gravity field applying the following equation:

\begin{equation}
   \Bigg[\dfrac{B_{pos}}{G} \Bigg] = \sqrt{\frac{sin \psi}{sin \alpha} \left(\nabla P + \rho \bigtriangledown \phi \right) . 4\pi R} = \sqrt{\Sigma_B \left(\nabla P + \rho \nabla \phi \right)  4\pi R},
   \label{eq:bfieldKoch}
\end{equation}

where density ($\rho$) is in g cm$^{-3}$, gravitational potential gradient ($\nabla \phi$) is in cm s$^{-2}$ , and B-field curvature radius (R) is in cm, resulting B-field in Gauss. Here, $\Sigma_B$ is the magnetic field significance shown in Section \ref{sec:methods}. For simplicity, we neglected the pressure gradient ($\nabla P$) which is typically small compared to gravity potential.

Although we have restricted our calculations to an area of 0.15~pc of radius around the emission peak, 
the gravitational potential at each point within this area is generated considering the contribution of all points on the map. The curvature C is computed from Eq. \ref{eq:curvature} \citep[][see q. 6]{Koch2012}:

\begin{equation}
    \label{eq:curvature}
    C \equiv \dfrac{1}{R} = \dfrac{2}{d} \cos \left(\dfrac{1}{2} \left[\pi - \Delta PA \right] \right),
\end{equation}

where d is the separation between two neighboring magnetic field segments and $\Delta$ PA is the difference between their position angles.
The magnetic field strength is obtained using the average values of the volume density (see Table \ref{tab:polresults}) over the area enclosed within 0.15~pc from the Hub-N center, together with $\Sigma_B$ (0.7), and the field radius (1.3$\times10^{18}$ cm).
%\LEt{Please check that I have retained your intended meaning.}

In Table \ref{tab:polresults} we present the estimation of the magnetic field strength in Hub-N obtained with this method, finding $\sim$0.6~mG. This value is consistent with the magnetic field estimated through the DCF method ($\sim$0.8~mG, Sec \ref{sec:methodI}). Regarding Hub-S, as we explain in the previous section, the present polarization data do not allow us to perform a similar study.

\begin{table*}[!t]
    \centering
    \small
    \caption{Physical properties of the two hubs 
    } 
    \label{tab:polresults}
    \begin{threeparttable}[b]
    \begin{tabular}{c c c c c c c c c c c c c}
    \hline\hline
        Source & $N_{\mathrm{mm}}$ \tnote{a} & & $M_{\mathrm{0.15pc}}$ \tnote{b} & $n_{\mathrm{0.15pc}}$ \tnote{b} & $\sigma_{\mathrm{1D,nth}}$ \tnote{c} & 
        $\langle$ PA $\rangle$ \tnote{d} & $\sigma_{PA}$ \tnote{d} & $B_{\mathrm{DCF}}$ \tnote{e} & $B_{\mathrm{DCF}}^{\mathrm{mod}}$ \tnote{f} & $B_{\mathrm{P-IG}}$ \tnote{g}& $\lambda_{\mathrm{DCF/mod/P-IG}}$\tnote{h}& $\mathcal{M}_{\rm{A-DCF/mod/P-IG}}$ \tnote{h}\\
        & &  & (\msun) & ($\times 10^5$cm$^{-3}$) & (\kms) & ($^{\circ}$) & ($^{\circ}$) & (mG) & (mG) & (mG)\\
        \hline
         Hub-N & 4 &  & 126 & 1.3 & $0.83\pm0.04$ & $\sim$ 102 & $\sim$ 7.8 & $\sim0.8$%\pm0.6$ 
         & $\sim$ 0.7
         %\pm$0.1  
         & $\sim$ 0.6 & 0.3 / 0.4 / 0.4
         %$\pm$0.3 
         & 0.3 / 0.4 / 0.4
         %$\pm$0.3
         \\
         Hub-S & 13 &  & 105 & 1.1 & $0.88\pm0.02$ & $\sim$ 84 & $\sim$ 29 & $\sim0.2$
         %\pm0.04$ 
         & $\sim0.1$ & -- & 1.1 / 1.6
         %$\pm$2.4 
         & 1.1 / 1.6
         %$\pm$0.5
         \\
         \hline
    \end{tabular}
    \begin{tablenotes}
    \item[a] {$N_{\mathrm{mm}}$ is the number of fragments detected at 1.3~mm within a field of view of 0.3~pc of diameter \citep{Busquet2016}.}
    \item[b]
    {Mass and average density  within a radius of 0.15~pc inferred by modeling the SED and the radial intensity profile \citep{Busquet2016}.}
    \item[c] {Nonthermal turbulent component of the velocity dispersion obtained from NH$_3$\,(1,1) data, averaged over the 0.15~pc area. \citep{Busquet2013,Busquet2016}.}
 
    \item[d]{Average magnetic field position angle and dispersion computed as the standard deviation.
    }
    \item[e]{Magnetic field in the plane of the sky estimated using the DCF method.}
    \item[f]{Magnetic field in the plane of the sky estimated using the modified DCF method  \citep{Heitsch2001}.}
    \item[g]{Magnetic field in the plane of the sky estimated using the polarization--intensity gradient method \citep{Koch2012}} 
    \item[h]{Mass-to-magnetic-flux ratio and Alfvén Mach number estimated using $B_{\mathrm{DCF}}$,  $B_{\mathrm{DCF}}^{\mathrm{mod}}$, and
    $B_{\mathrm{P-IG}}$}
    \end{tablenotes}
    \end{threeparttable}
\end{table*}
    
\subsection{Mass-to-flux ratio and turbulence}
\label{sec:masstoflux}

The relevance of the magnetic force with respect to gravitational force can also be estimated by the mass-to-magnetic-flux ratio, stated in units of the critical value, 

\begin{equation}
    \lambda = \frac{(M/\Phi)_{\mathrm{observed}}}{(M/\Phi)_{\mathrm{crit}}} = 2.6 \times 10^{-21} \Bigg[\frac{N(\text{H}_{2})}{\text{cm}^{-2}}\Bigg]\Bigg[\frac{B_{\text{pos}}}{\mu\,G}\Bigg]^{-1}, 
    \label{eq:lambda}
\end{equation}

\noindent
where we have applied a correction of one-third, which considers the statistical projection effects \citep[see][]{Crutcher2004}.

In order to compare the role of the magnetic field between the two cores and to compare different magnetic field strength estimation as well, we used different $B_{\mathrm{pos}}$ values derived from both the standard DCF, the modified DCF relation, and the polarization--intensity gradient technique. We used the column density estimated by \cite{Busquet2016} within the 0.15~pc radius (N(H$_2$)= 1.1 or 0.9 $\times$  10$^{23}$ cm$^{-2}$ for Hub-N or Hub-S, respectively)

The mass-to-flux ratio is calculated for the three different estimates of the magnetic field strength given in Table~\ref{tab:polresults}. For Hub-N we obtain a value for $\lambda$ of between 0.3 and 0.4 (Table~\ref{tab:polresults}).   For this hub, we can also calculate the mass-to-flux ratio using the polarization--intensity gradient technique, which is totally independent of the estimation of the magnetic field. From Eq.~8 in \cite{Koch2012b}:

\begin{equation}
    \left( \dfrac{M}{\Phi} \right)_{norm} = \left\langle \Sigma_B ^{-1/2} \right \rangle \left(\dfrac{R}{R_o}\right)^{-3/2} \pi^{-1/2}  
,\end{equation}
where $R_0$=$R$ is the cloud radius.

Through this method we find a value of $\sim$ 0.7 which is about twice the value found previously, but still subcritical, that is, the magnetic field can hold the gravitational collapse. For Hub-S, the resulting mass-to-flux ratio from Eq.~\ref{eq:lambda} is between 1.1 and 1.6, which three to four times higher than in Hub-N. Hub-S is therefore magnetically supercritical.
Regarding the absolute value of $\lambda$, this is probably underestimated because the magnetic field average along the line of sight is not taken into account, which is probably more severe around the hub center \citep[\eg][]{Houde2009}. Indeed, the Hub-N $\Sigma_B$ map shows values lower than 1 in most of the hub. In addition, both hubs show clear signs of active and  ongoing star formation revealed by the presence of H$_2$O and CH$_3$OH masers \citep{Jaffe1981,Wang2006,Sugiyama2017}, infrared sources \citep{Povich2016}, and molecular outflow activity (Busquet, private communication). This is an indication that the two hubs are supercritical.

Finally, to assess the importance of turbulence with respect to the magnetic field, we estimated the Alfv\'en Mach number,  $\mathcal{M}_{\rm{A}}$, expressed as $\mathcal{M}_{\rm{A}}=\sqrt{3}\sigma_\mathrm{1D,nth}/v_{\mathrm{A}}$, where $v_{\mathrm{A}}=B_{\mathrm{tot}}/\sqrt{4\pi\rho}$ is the Alfv\'en speed. The uncertainty in the Alfvén Mach number is the same as the uncertainty in $\sigma_{\mathrm{P.A.}}$. As can be seen in Table~\ref{tab:polresults},  $\mathcal{M}_{\rm{A}}<1$ in Hub-N, indicating sub-Alfvénic conditions and therefore that magnetic energy dominates over turbulence. Similar results have been found in other IRDCs \citep[\eg][]{Pillai2015,Soam2019}. On the contrary,  $\mathcal{M}_{\rm{A}}>1$ in Hub-S (\ie\ super-Alfvénic conditions), and therefore turbulence could play a more important role in this region.

%%%%%%%%%%%%%%%%%%%%%%%%%%%%%%%%%%%%%%%DISCUSSION %%%%%%%%%%%%%%%%%%%%%%%%%%%%%%%%%%%%%%%%%%%%%%%%%%%

\section{Discussion}\label{s:discuss}

With the present study we show that the magnetic field orientation in Hub-N is relatively uniform along the east--west direction, being perpendicular to the major axis of the hub--filament system. 
Overall, based on the polarization--intensity gradient technique, we conclude that the magnetic field probably cannot prevent gravitational collapse,
except in the north--south ridge where $\Sigma_{B} > 1$ in some positions. 
Hub-S, on the other hand, presents a bimodal magnetic field distribution, in which each component converges toward each local minimum identified with the intensity gradient map. At larger distances from the hub, the magnetic field appears perpendicular to the dense filaments.

Our aim is to investigate the relevance of the magnetic field in hub--filament systems and the fragmentation processes that they undergo to form star clusters. In this section, we first discuss our magnetic field strength estimation, which was derived using three different methods, and the role of the magnetic field at different scales, from the large-scale hub--filament system down to the 0.1~pc scale. We use this information to assess the role of the magnetic field in producing different fragmentation levels in these two hubs.

\subsection{Magnetic field strength estimation}

In this work we present three different estimations of the magnetic field strength based on the DCF \citep{DavisPhysRev, ChandrasekharFermi1953}, the modified DCF method presented by \cite{Heitsch2001}, and the polarization--intensity gradient technique \citep{Koch2012}.

In Hub-N, the three estimates of the magnetic field lead to consistent results, with values of between $0.6$ and $0.8$~mG. The largest value is provided by DCF, as you would expect from an upper limit of the magnetic field (see Section \ref{sec:methodI}).
%which is \textcolor[rgb]{0.984314,0.00784314,0.027451}{consistent with its description as upper limit we give in} \LEt{ the meaing here is unclear; please consider rewording or expanding slightly.} Section \ref{sec:methodI}.

To compare the magnetic field in both Hubs we use the DCF and modified DCF methods. Unlike the results found by \cite{Busquet2016}, who found similar values for the magnetic field in both hubs, we find a stronger magnetic field in Hub-N than in Hub-S by a factor of approximately equal to or less than four.
%\LEt{ the meaning of leaving simply "less than four" is unclear; that is, it appears extremely vague.}. 
It is important to note that we only considered an area of 0.15~pc radius around the peak emission in both hubs. Therefore, we restrict the calculation to the area where we have more detections, especially in Hub-S.

\subsection{Gravity and magnetic field forces}

Recent results obtained with the \textit{Planck} satellite have shown that magnetic fields in the solar-neighborhood molecular clouds tend to be parallel to low-density structures and perpendicular to high-column-density structures \citep{PlanckColl2016,Soler2019}. Numerical simulations under the scenario of global, hierarchical gravitational collapse \citep[\eg][]{GomezandVazque-Semadeni2014,Vazquez-Semadeni2019} predict that filaments gain their mass from their surrounding material, which flows down %\LEt{Please check that I have retained your intended meaning.}
the gravitational potential (\ie\ a clump or hub). The magnetic field is dragged by the collapsing material, following the accretion flow that feeds the filament, which is why the magnetic field tends to be oriented mostly perpendicular to the filaments \citep{Gomez2018}. Within dense filaments, \cite{Gomez2018} point out that the magnetic field lines would bend into the filament because of the flow toward the hub, which shows a characteristic `U' shape.
Such longitudinal accretion flows along filaments have been observed in several filamentary regions 
(\eg\ \citealt[][in Serpens South]{Kirk2013,Fernandez-Lopez2014}, \citealt[in SDC\,13]{Peretto2014}, \citealt[][in eight filamentary high-mass star-forming clouds]{Lu2018}), while magnetic fields oriented parallel to  dense filaments have recently been observed in a few regions \citep{Monsch2018,Liu2018, Pillai2020}  .

In this work, we assume that the intensity gradient of the 350~\mum emission is an approximation of the inertial term in the MHD equation (left-hand term in Eq. \ref{eq:MHD}) which can trace the flow of matter.
Based on the kinematics of the N$_2$H$^{+}$\,(1--0) line, \citet{Chen2019} find large-scale velocity gradients along the filament, $\sim0.5$~\kms\,pc$^{-1}$, indicative of inflow motions toward the two hubs that harbor deeply embedded protoclusters, with a mass accretion rate along the filaments of  $\sim{10}^{-4}\,{M}_{\odot }\,{\mathrm{yr}}^{-1}$.
Figure~\ref{fig:deltamap_North} (right panel) shows the relative orientation between the
% \LEt{Please spell out all acronyms the \uline{first} time they appear in the paper, followed by the abbreviation in parentheses, both in the abstract and again in the main text. After that, please\uline{ only} use the abbreviation. See A and A language guide Section 5.2.4 www.aanda.org/language-editing}
position angles of the magnetic field and the intensity gradient in Hub-N. At large distances from the hub--filament system, the magnetic field is roughly parallel to the intensity gradient ($|\delta|\simeq0$), becoming perpendicular as it enters the densest parts of both filament and hub.
This situation is consistent with a scenario in which filaments are accreting gas material from their surrounding environment and the flow of matter onto the filament drags the magnetic field. %\textcolor[rgb]{0.984314,0.00784314,0.027451}{due to the} collapsing gas.
%\LEt{ the meaning here is unclear; specifically the relationship with the collapsing gas; please consider rewording here for clarity.} 
Indeed, both east and west of Hub-N and F10-E, the gravity force dominates over magnetic fields (see Fig.~\ref{fig:sigmaB_North}), in agreement with numerical simulations under the scenario of GHC \citep[\eg][]{Gomez2018,Vazquez-Semadeni2019}.

Despite the presence of inflow motions within the two filaments converging toward Hub-N \citep[][]{Chen2019}, the magnetic field in the dense regions appears to be perpendicular to the intensity gradient (\ie\ perpendicular to the gas flow). 
From Fig.~\ref{fig:fil-spine} (upper panel) it is clear that the magnetic field in the dense filament does not significantly change its orientation. This might be because the CSO observations lack the necessary resolution to fully resolve the spine of the filaments seen in N$_2$H$^+$
and could be indicative of an underlying complex B-field morphology.
There is a slight tendency of the magnetic field to appear distorted, although we cannot distinguish the U-shaped magnetic field structure predicted by \citet{Gomez2018}. 
This could be explained by the lack of resolution in our CSO 350~\mum observations, where $10''$ at 1.98~kpc  corresponds to $\sim0.1$~pc, whereas the numerical simulations of \citet{Gomez2018} were performed with a spatial resolution of $1.2\times10^{-4}$~pc. Under this hypothesis, the magnetic field is expected to be more complex at smaller scales.

Another plausible explanation for the lack of the U-shaped magnetic field 
%\LEt{ please expand slightly here for clarity; i.e. "...explanation for this distortion could be...".} 
could be found in terms of optical depth effects. The 350~\mum polarization data might trace an upper layer and might not penetrate equally as deep as N$_2$H$^+$ and NH$_3$.
In addition, simulations show that this U-shaped structure is more pronounced as density increases. It is interesting to mention that in the simulations of \citet{Gomez2018} the accretion flow within filaments takes place where the magnetic field is weaker. 
Our results from Fig.~\ref{fig:sigmaB_North} suggest that the magnetic field seems to dominate over gravity along a north--south ridge encompassing the CSO 350~\mum dust emission peak that harbors the primary dust core detected at high angular resolution \citep{Busquet2016,Ohashi2016}. This could indicate that the gas flow follows the magnetic field, or, alternatively, that the magnetic field structure inside filaments (\ie\ at smaller scales) is much more complex \citep[\eg][]{Li2019}. Unfortunately, the current angular resolution does not allow us to disentangle the magnetic field structure within the dense filament. Therefore, dust polarization observations at higher angular resolution (e.g. ALMA) are required to further investigate the magnetic field structure within dense filaments and cores and its specific role in channeling material from the dense filament toward the embedded protocluster.

It is interesting to mention that in Fig.~\ref{fig:sigmaB_North}, and especially in Fig.~\ref{fig:deltamap_North}, the region in which the relative orientation of the intensity gradient and the magnetic field is roughly parallel (i.e., |$\delta| \simeq$ 0) seems to converge towards the position of the central dust continuum core MM1a \citep{Busquet2016}. This supports the idea that MM1a is accreting material from the environment east and west of the central core. In addition, as indicated by the velocity gradients detected along the filament's longer axis
\citep{Chen2019}, gas flows through the filament to the hub. Therefore, in Hub-N we have large-scale accretion from the east and west and an accretion flow through the filament to the hub in the south to north direction.

\subsection{Fragmentation and magnetic field properties toward the hubs}

In a previous study, \cite{Busquet2016} found that the two hubs in the IRDC G14.2 present different levels of fragmentation. By extrapolating the magnetic field strength obtained at infrared wavelengths that trace the diffuse gas to the dense hubs harboring a deeply embedded protocluster, \citet{Busquet2016} obtain similar values of the magnetic field, mass-to-flux ratio, and Alfvén Mach number in both hubs.
Now, thanks to the submillimeter CSO polarization data we can penetrate deeper into the hub--filament system. 
As shown in Section 4.3, we find some differences in the magnetic field strength between both hubs, namely greater strength toward Hub-N. As a consequence, the mass-to-flux ratio and the Alfvénic Mach number are 
also different. While Hub-N seems to have sub-Alfvénic conditions, Hub-S presents super-Alfvénic conditions, indicating that turbulence plays a greater role than the magnetic field, at least at the scales traced by the CSO observations ($\sim$0.1~pc). These conditions are different from those inferred by \citet{Busquet2016}, in which the magnetic field was estimated to be of the same order as in Hub-N and to display sub-Alfvénic conditions.

\cite{Palau2013} investigate the fragmentation of massive dense cores down to $\sim1000$~au by comparing high-angular-resolution and high-sensitivity observations with the radiation magneto-hydrodynamic simulations of \cite{Commercon2011}. In this work, the highly magnetized cores show low fragmentation while the weakly magnetized cores present a higher fragmentation level.  Recently, \cite{Fontani2016,Fontani2018} find that fragmentation is inhibited when the initial turbulence is low (sonic Mach number $\mathcal{M}\sim3$), independently of the other physical parameters. In addition, these latter authors point out that a filamentary distribution of the fragments is favored in a highly magnetized clump, while other morphologies are also possible in a weaker magnetic field scenario. In any case, \citet{Fontani2016} show that the weakly magnetized simulations display more fragments. 
In the two hubs of IRC G14.2, the different levels of fragmentation could be explained by the different values of magnetic field strength. Both hubs share some physical properties such as the internal structure of the envelope (temperature and density distribution) where the fragments are embedded, and similar Mach number, $\mathcal{M}\simeq6$ \citep{Busquet2016}. 
The lower fragmentation level observed toward Hub-N, together with the homogeneous and well aligned magnetic field structure, the derived values of mass-to-flux ratio, and $\mathcal{M_{\mathrm{A}}}$, favors a scenario in which the magnetic field plays an important role in regulating the collapse and fragmentation processes in these two hubs, possibly by slowing down the star formation process. 

In a very recent work, \cite{Palau2020} estimate the magnetic field strength and fragmentation level in a sample of massive dense cores, and find that fragmentation level seems to correlate with core density, although with significant scatter, while there is no correlation with magnetic field strength. The correlation of fragmentation level with density, which is consistent with previous works \citep[\eg][]{Gutermuth2011,Palau2014,Mercimek2017,Sanhueza2019}, suggests that gravity plays an important role in the fragmentation process. However, for those cores with similar densities in the Palau et al. (2020) sample, the magnetic field strength is greater in the cores with lower fragmentation, and the different magnetic field strengths could explain the scatter in the fragmentation-versus-density relation. Thus, the magnetic field seems to have a modulating effect to the dominant role of gravity. The work presented here is fully consistent with the results of Palau et al. (2020), as the two hubs in G14 present similar densities and the magnetic field seems to be larger for Hub-N where fragmentation is lower.

Finally, it is interesting to look at the immediate surroundings of these two hubs in order to investigate whether or not the environment could lead to different levels of fragmentation. On the one hand,  the \ion{H}{II} region IRAS\,18153$-$1651 lies in the vicinity of
Hub-N, with a bolometric luminosity of $\sim10^4$~\lsun\ \citep{Jaffe1982}. Recent optical spectroscopic observations reveal that this IRAS source consists of two main sequence stars of spectral type B1 and B3 and that the \ion{H}{II} indeed contains a recently formed star cluster \citep{Gvaramadze2017}. Moreover, \cite{Gvaramadze2017} report the discovery of an optical arc nebula located at the western edge of the IRAS source, which was interpreted as the edge of the photoionzing wind bubble blown by the B1 star.  This one-sided appearance could result from the interaction between the bubble and the photoevaporation flow from the molecular gas associated with Hub-N and its associated filaments \citep{Gvaramadze2017}. 
The presence of this wind bubble powered by the 
\ion{H}{II} region could compress the gas and produce the observed uniformity of the magnetic field, similarly to the Pipe Nebula where the highly magnetized regions with a uniform magnetic field arise from gas compression due to the collision of filaments \citep{Frau2015}, consistent with the findings of \citet{Busquet2013}, where higher temperatures were measured at the eastern edge of Hub-N.

Regarding the dense gas surrounding the hubs, both hubs display a network of filaments that seem to converge into the hubs. However, in the northern part, the hub--filament system consists of a dominant filament in the north--south direction with a clear velocity gradient along the filament's longer axis that transports material into Hub-N \citep{Chen2019}. Hub-S on the other hand presents a more complex network of filaments, with one filament coming from the east and another filament that approaches the northern part of the hub from the southwest with redshifted velocities \citep{Busquet2013,Chen2019}. 
We cannot discard that such a distribution is reflecting a projection effect, 
where the more complex filament distribution in the south \citep{Busquet2016,Chen2019} is smearing out the magnetic field observed by an averaging effect at the current CSO scale.
Clearly, this scenario should be further investigated through interferometric observations (SMA, ALMA) with higher sensitivity polarization data to trace the magnetic field around the whole IRDC complex.

\section{Summary and Conclusions}\label{s:concl}

We present the results of CSO 350~\mum dust polarization observations toward Hub-N and Hub-S in the IRDC G14.225-0.506 and an analysis of the polarization–intensity gradient using the method developed by \citet{Koch2012}.
The main findings of this work can be summarized as  follows.

\begin{enumerate}

\item Toward Hub-N, we find an almost uniform
magnetic field orientation in the east--west direction, which is nearly perpendicular to the major axis of the hub-filament system. However, in Hub-S the magnetic field presents a bimodal distribution, with one component going from the northeast to the southwest and the other component going from the northwest to the southeast. 
     
\item The intensity gradient in Hub-N presents a single local minimum coinciding with the bright dust continuum core MM1a, and the prevalent east--west magnetic field orientation slightly deviates when approaching the dust core. In Hub-S, the intensity gradient reveals two minima, reflecting the bimodal distribution of the magnetic field, as each component points toward one of the two intensity gradient minima. This result suggests a scenario in which the magnetic field is dragged by the collapsing cores. 

\item The analysis of the |$\delta$|  and $\Sigma_B$ maps in Hub-N indicates that, near the hubs, gravity dominates the magnetic field. The intensity gradient is parallel to the magnetic field lines both east and west of the hub, suggesting that the magnetic field is channeling material toward the central ridge or that the B-field is being dragged by gravity.
We find that magnetic field segments do not change their direction into the filament. However, this could be due to the relatively coarse CSO resolution which falls short of resolving the B-field morphology in the close vicinity around the spine of the filaments.
Combined with the velocity information from previous works, in Hub-N we see large-scale accretion from the east and west toward the central dust core and the accretion flow within the main filament (in the south--north direction) that feeds the hub.

\item We find higher values of the magnetic field strength in Hub-N than in Hub-S. This supports the idea that the different levels of fragmentation in these two hubs could result from differences in the magnetic field. However, we do not find an extreme difference between the magnetic field characteristics of the two hubs. Therefore, key differences could be found in the environment, perhaps in the influence of the nearby \ion{H}{II} region in Hub-N or in the more complex distribution of filaments in Hub-S. 
\end{enumerate}

In summary, our analysis of the magnetic field in the two hubs of the IRDC G14.2 reveals that the magnetic field could play a role in the fragmentation process and in the formation and evolution of dense filaments. Further observations of dust polarization with higher sensitivity and at higher angular resolution are needed to further investigate the magnetic field structure within the dense filaments and at core scales.

\begin{acknowledgements}
We are grateful to Huei-Ru Vivien Chen for sharing with us her studies on the region.
This material is based upon work at the Caltech Submillimeter Observatory, which is operated by the California Institute of Technology. 
N.AL. acknowledges support from Ministery of Science and Innovation fellowship for predoctoral contracts in Spain and Ministery of Science and Technology (MoST) Summer Program in Taiwan for Spanish Graduate Student. 
N.AL., G.B., and J.M.G. are supported by the Spanish grant AYA2017-84390-C2-R (AEI/FEDER, UE). 
P. M. K. acknowledges support from Ministery of Science and Technology (MoST) grants MOST 108-2112-M-001-012 and MOST 109-2112-M-001-022 in Taiwan, and from an Academia Sinica Career Development Award. 
H.B.L. is supported by the Ministry of Science and Technology (MoST) of Taiwan (Grant MOST. 108-2112-M-001- 002-MY3).
P.T.P.H. is supported by Ministry of Science and Technology (MoST) of Taiwan Grant MOST 108-2112-M-001-016-MY1.
A.P. acknowledges financial support from CONACyT and UNAM-PAPIIT IN113119 grant, M\'exico.

\end{acknowledgements}

\bibliographystyle{aa} 
\bibliography{bibliografia}

\begin{appendix}

\section{Observations details} \label{sec:apendixA}

\begin{figure}[!ht]
    \centering
    \includegraphics[width=0.5\textwidth]{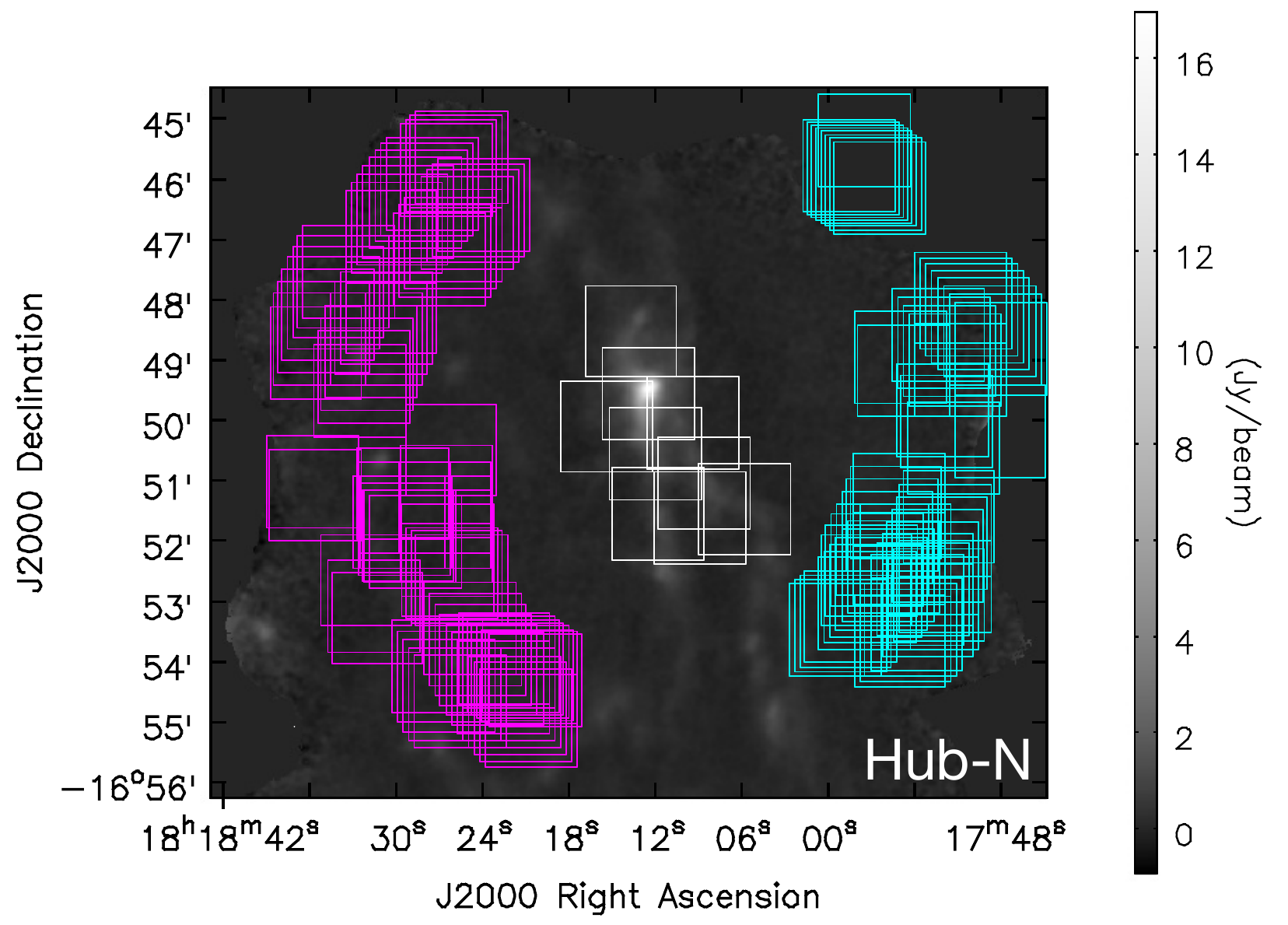}
    \includegraphics[width=0.5\textwidth]{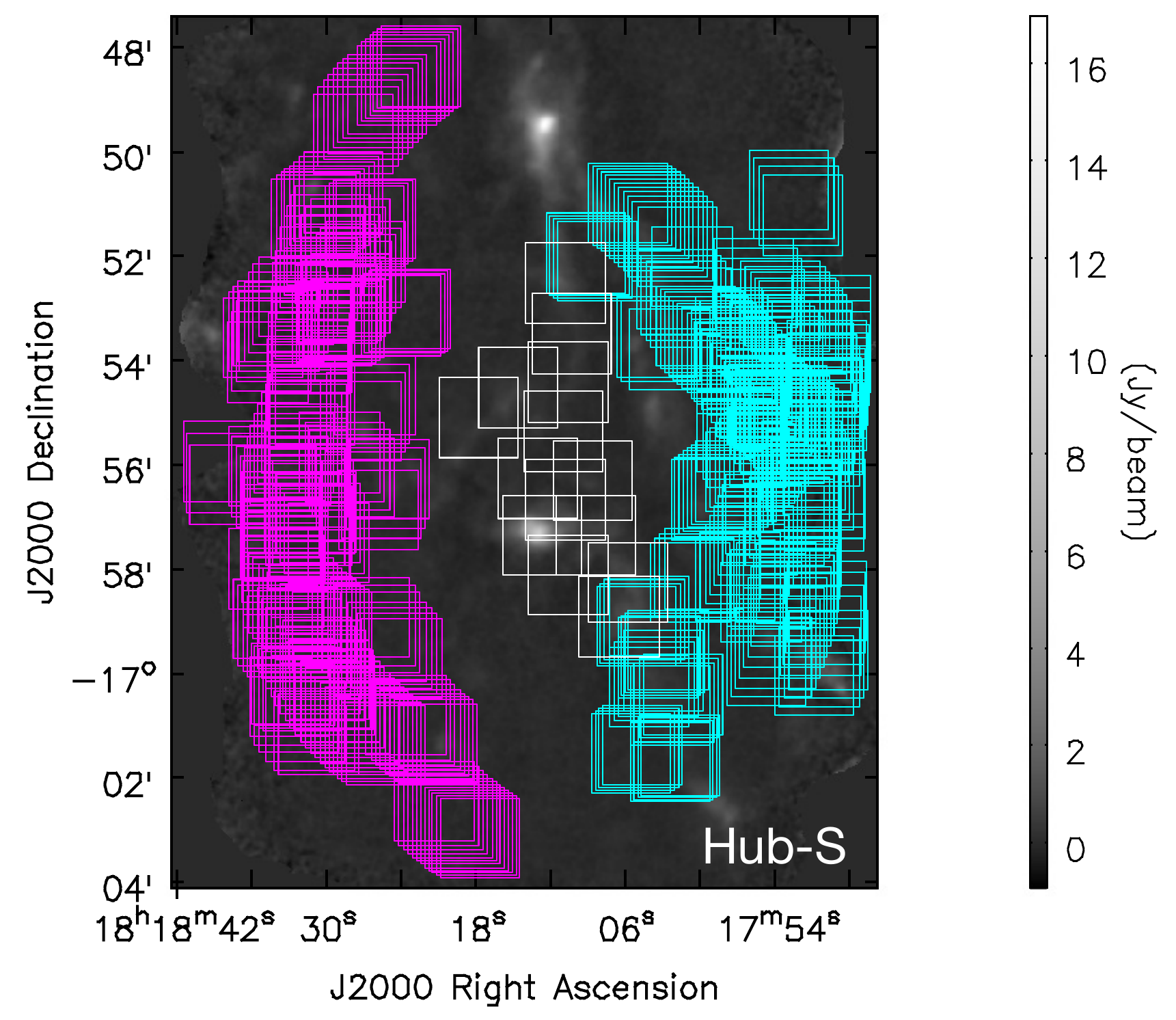}
    \caption{Pointings for SHARP observations. The white squares%\LEt{Please check that I have retained your intended meaning.}
     depict the fields of view that make up the mosaic around the source (corresponding to Table \ref{tab:pointings}). Magenta and cyan squares show the off-beam positions. Upper and lower panels represents pointings toward Hub-N and Hub-S, respectively.}
    \label{fig:regions}
\end{figure}

In this section we expand the information on the CSO observations. Figure~\ref{fig:regions} shows the different pointings conducted toward the science target (in white), and off-beam positions (in magenta and cyan) for both Hub-N (upper panel) and Hub-S (bottom panel). The pointing coordinates of the science target are listed in
Table~\ref{tab:pointings}.

\begin{table}
\caption{Pointings of the CSO SHARP observations in G14.2\label{tab:pointings}}
\begin{tabular}{ c c c }
\hline\hline
  ID & R.A. & Decl. \\
  & (J2000) & (J2000) \\
  \hline
  Hub N && \\\hline
g14 n1 & 18\rah18\ram13\ras630  &       -16\decd48\decm30\decs40  \\
g14 n2 &                18\rah18\ram12\ras480    &       -16\decd49\decm31\decs90  \\
g14 n3 &                18\rah18\ram12\ras060    &       -16\decd50\decm31\decs90  \\
g14 n4 &                18\rah18\ram11\ras697    &       -16\decd51\decm31\decs90  \\
g14 n5 &                18\rah18\ram08\ras615    &       -16\decd51\decm01\decs50  \\
g14 n6 &                18\rah18\ram08\ras980    &       -16\decd51\decm36\decs00  \\
g14 n7 &                18\rah18\ram09\ras451    &       -16\decd50\decm01\decs50  \\
g14 n8 &                18\rah18\ram15\ras302    &       -16\decd50\decm04\decs50  \\
g14 n9 &                18\rah18\ram10\ras705    &       -16\decd48\decm48\decs00  \\
g14 n10 &               18\rah18\ram05\ras689    &       -16\decd51\decm26\decs99  \\\hline
  Hub S \\\hline
g14 s1 &                18\rah18\ram17\ras080    &       -16\decd57\decm19\decs90  \\
g14 s2 &                18\rah18\ram12\ras790    &       -16\decd57\decm19\decs90  \\
g14 s3 &                18\rah18\ram08\ras450    &       -16\decd57\decm19\decs90  \\
g14 s4 &                18\rah18\ram05\ras783    &       -16\decd58\decm13\decs81  \\
g14 s5 &                18\rah18\ram10\ras632    &       -16\decd58\decm05\decs12  \\
g14 s6 &                18\rah18\ram10\ras875    &       -16\decd56\decm38\decs16  \\
g14 s7 &                18\rah18\ram06\ras554    &       -16\decd58\decm53\decs42  \\
g14 s8 &                18\rah18\ram15\ras238    &       -16\decd56\decm38\decs16  \\
g14 s9 &                18\rah18\ram18\ras996    &       -16\decd56\decm43\decs48  \\
g14 s10 &               18\rah18\ram13\ras056    &       -16\decd56\decm13\decs81  \\
g14 s11 &               18\rah18\ram17\ras541    &       -16\decd55\decm56\decs42  \\
g14 s12 &               18\rah18\ram14\ras632    &       -16\decd58\decm01\decs65  \\
g14 s13 &               18\rah18\ram20\ras571    &       -16\decd56\decm06\decs85  \\
g14 s14 &               18\rah18\ram14\ras719    &       -16\decd55\decm27\decs92  \\
g14 s15 &               18\rah18\ram17\ras645    &       -16\decd55\decm03\decs92  \\
g14 s16 &               18\rah18\ram14\ras719    &       -16\decd54\decm29\decs42  \\
g14 s17 &               18\rah18\ram10\ras956    &       -16\decd55\decm20\decs42  \\
g14 s18 &               18\rah18\ram10\ras642    &       -16\decd54\decm23\decs42  \\
g14 s19 &               18\rah18\ram10\ras329    &       -16\decd53\decm27\decs92  \\
g14 s20 &               18\rah18\ram10\ras747    &       -16\decd52\decm29\decs42  \\
g14 s21 &               18\rah18\ram08\ras760    &       -16\decd56\decm15\decs92  \\\hline
\end{tabular}
\end{table}

\end{appendix}

\end{document}